\documentclass[fleqn,usenatbib]{mnras}

\usepackage[T1]{fontenc}

\DeclareRobustCommand{\VAN}[3]{#2}
\let\VANthebibliography\thebibliography
\def\thebibliography{\DeclareRobustCommand{\VAN}[3]{##3}\VANthebibliography}

\usepackage{graphicx}	% Including figure files
\usepackage{amsmath}	% Advanced maths commands
\usepackage{amssymb}	% Extra maths symbols
\usepackage{bm}
\usepackage{xcolor}
\usepackage{caption}

\usepackage{newtxtext,newtxmath}
\title[Inhomogeneous UVLF During Reionization]{Cosmic Variance and the Inhomogeneous UV Luminosity Function of Galaxies During Reionization}

\author[T. Dawoodbhoy et al.]{
Taha Dawoodbhoy,$^{1}$\thanks{E-mail: tahad@astro.as.utexas.edu}
Paul R. Shapiro,$^{1}$
Pierre Ocvirk,$^{2}$
Joseph S. W. Lewis,$^{3,4}$
Dominique Aubert,$^{2}$
\newauthor
Jenny G. Sorce,$^{5,6,7}$
Kyungjin Ahn,$^{8}$
Ilian T. Iliev,$^{9}$
Hyunbae Park,$^{10,11,12}$
Romain Teyssier,$^{13}$
\newauthor
Gustavo Yepes$^{14,15}$
\\
$^{1}$Department of Astronomy and Texas Center for Cosmology and Astroparticle Physics, The University of Texas at Austin, Austin, TX 78712-1083, USA\\
$^{2}$Université de Strasbourg, Observatoire astronomique de Strasbourg, UMR 7550, F-67000 Strasbourg, France\\
$^{3}$Zentrum f\"ur Astronomie der Universit\"at Heidelberg, Institut f\"ur Theoretische Astrophysik, Albert-Ueberle-Stra\ss e 2, 69120 Heidelberg, Germany\\
$^{4}$Max-Planck-Institut f\"ur Astronomie, K\"onigstuhl 17, D-69117 Heidelberg, Germany\\
$^{5}$Univ. Lille, CNRS, Centrale Lille, UMR 9189 CRIStAL, F-59000 Lille, France\\
$^{6}$Leibniz-Institut für Astrophysik Potsdam (AIP), An der Sternwarte 16, D-14482 Potsdam, Germany\\
$^{7}$Universit\'e Paris-Saclay, CNRS, Institut d'Astrophysique Spatiale, 91405, Orsay, France\\
$^{8}$Chosun University, 375 Seosuk-dong, Dong-gu, Gwangjiu 501-759, Korea\\
$^{9}$Astronomy Center, Department of Physics \& Astronomy, Pevensey II Building, University of Sussex, Falmer, Brighton BN1 9QH, United Kingdom\\
$^{10}$Lawrence Berkeley National Laboratory, CA 94720-8139, USA \\
$^{11}$Berkeley Center for Cosmological Physics, UC Berkeley, CA 94720, USA \\
$^{12}$Kavli IPMU (WPI), UTIAS, The University of Tokyo, Kashiwa, Chiba 277-8583, Japan \\
$^{13}$Department of Astrophysical Sciences, Princeton University, Princeton, NJ 08540, USA\\
$^{14}$Departamento de F\'isica Te\'orica M-8, Universidad Aut\'onoma de Madrid, Cantoblanco, 28049, Madrid, Spain\\
$^{15}$Centro de Investigaci\'on Avanzada en F\'isica  Fundamental (CIAFF), Universidad Aut\'onoma de Madrid, 28049 Madrid, Spain
}

\date{Accepted XXX. Received YYY; in original form ZZZ}

\pubyear{2022}

\begin{document}
\label{firstpage}
\pagerange{\pageref{firstpage}--\pageref{lastpage}}
\maketitle

% Abstract of the paper
\begin{abstract}
When the first galaxies formed and starlight escaped into the intergalactic medium to reionize it, galaxy formation and reionization were both highly inhomogeneous in time and space, and fully-coupled by mutual feedback. To show how this imprinted the UV luminosity function (UVLF) of reionization-era galaxies, we use our large-scale, radiation-hydrodynamics simulation CoDa II to derive the time- and space-varying
halo mass function and UVLF, from $z\simeq6$--15.
That UVLF  correlates strongly with local reionization redshift: earlier-reionizing regions have UVLFs that are higher, more extended to brighter magnitudes, and flatter at the faint end than later-reionizing regions observed at the same $z$.
In general, as a region reionizes, the faint-end slope of its local UVLF flattens, and, by $z=6$ (when reionization ended), the global UVLF, too, exhibits a flattened faint-end slope, `rolling-over' at $M_\text{UV}\gtrsim-17$.
CoDa II's UVLF is broadly consistent with cluster-lensed galaxy observations of the Hubble Frontier Fields at $z=6$--8, including the faint end, except for the faintest data point at $z=6$, based on one galaxy at  $M_\text{UV}=-12.5$.
According to CoDa~II, the probability of observing the latter is $\sim5\%$.
However, the effective volume searched at this magnitude is very small, and is thus subject to significant cosmic variance.
We find that previous methods adopted to calculate the uncertainty due to cosmic variance underestimated it on such small scales by a factor of 2--4, primarily by underestimating the variance in halo abundance when the sample volume is small.

\end{abstract}

% Select between one and six entries from the list of approved keywords.
% Don't make up new ones.

%%%%%%%%%%%%%%%%%%%%%%%%%%%%%%%%%%%%%%%%%%%%%%%%%%

%%%%%%%%%%%%%%%%% BODY OF PAPER %%%%%%%%%%%%%%%%%%

\section{Introduction}

Reionization -- the process by which starlight from early galaxies leaks into the surrounding intergalactic medium (IGM), gradually changing its ionization state from almost completely neutral before the first stars formed, at 
$z \gtrsim 20$, to almost completely ionized at $z\lesssim5.5$ -- was highly inhomogeneous in space and time \citep[see, e.g.,][]{YHO12,DF18,Bosman22}. 
The inhomogeneity was seeded by the gravitational growth of density perturbations in the early Universe, which, when non-linear, formed small overdense regions packed with clusters of dark matter halos in some places, and vast underdense voids in others.
The largest-amplitude perturbations formed these non-linear structures first, and so the most overdense regions were the earliest sites of star formation.
As such, these regions were the first to reionize, and were the origins of the first H II bubbles that grew radially outward from them, `exporting' their excess ionizing radiation to nearby lower-density regions and reionizing them, as well, in the process \citep[][]{Dawoodbhoy18}.
Over time, regions of progressively smaller-amplitude overdensity also reached their non-linear phase, forming more galaxies and stars, making existing H II bubbles larger and forming new ones where none had been before, eventually overlapping and filling all of space to complete the Epoch of Reionization
{  (``EOR'')}.
Thus, the local reionization history of any given region is strongly correlated with its local overdensity, and it is crucially important to take this correlation into account when making predictions for observables that depend on both the density and the ionization state of the observed region.
As we stress throughout this paper, the high-redshift UV luminosity function (UVLF; the number density of galaxies per unit UV luminosity or absolute magnitude, denoted $\Phi$) -- especially that for faint galaxies (absolute UV magnitudes $M_\text{UV} \gtrsim -16$) observed in small volumes through high-magnification gravitational lenses -- is one such observable { \citep[see, e.g.,][for a semi-analytic study]{KC11}}.

Recent analyses of high-$z$ galaxies found via the lensing clusters in the Hubble Frontier Fields (HFF) \citep{Lotz17} have led to some debate about the shape of the faint-end of the UVLF, especially at $z\sim6$.
The UVLF is often fit and parameterized with the Schechter function \citep{Schechter76}, which asymptotes to a power law at the faint end ($\Phi \propto (L/L_*)^\alpha$, $\alpha < 0$, for $L\ll L_*$) and an exponential cut-off at the bright end ($\Phi \propto \exp(-L/L_*)$, for $L\gg L_*$).
Some HFF studies \citep{Bouwens17,Atek18} have argued that the faint-end of the $z\sim6$ UVLF deviates from the power law behavior of the Schechter function by gradually flattening in slope faint-ward of $M_\text{UV} \sim -16$, while others \citep{Livermore17,Ishigaki18} have argued that the power law behavior is maintained well below this luminosity, with no evidence of a change in slope.
However, the observations and analyses of these extremely faint galaxies at high redshift are necessarily volume-limited; the galaxies can only be observed if they are located in particular regions such that their magnification by the foreground cluster is sufficient to raise their apparent brightness above the survey's flux limit.
These small-volume, high-magnification observations are subject to significant uncertainties, due to both potential errors in the lensing model and cosmic variance.
In this work, we seek to understand the latter.

There are several factors that cause the UVLF -- or the star formation rate, on which the UVLF strongly depends -- to vary from region to region \citep[see, e.g.,][]{Dawoodbhoy18}:
\begin{enumerate}
    \item overdense/early-reionizing regions have higher halo number densities than underdense/late-reionizing regions, especially at the high-mass end, and higher mass halos have higher star formation rates than lower mass halos;
    \item halos of a given mass in overdense/early-reionizing regions have higher star formation rates on average than halos of the same mass in underdense/late-reionizing regions;
    \item low-mass halos ($M \lesssim 10^{9.5}~\text{M}_\odot$) have lower star formation rates in regions that have already been reionized at a given redshift than regions that have yet to be reionized.
\end{enumerate}
Accurate modeling of these factors and their contribution to the cosmic variance of the UVLF requires the use of large-scale, high-resolution, fully-coupled radiation-hydrodynamics simulations, because they are highly contingent on the complex mutual feedback between galaxy formation and reionization.
The last factor, in particular, is due to the feedback of ionizing radiation photo-heating the IGM in the vicinity of low-mass halos, thereby inhibiting their ability to accrete gas and form stars \citep[cf.][]{SGB94}.
Since low-mass halos preferentially occupy the faint end of the UVLF, this reionization-induced suppression of star formation will have a significant impact on the faint end, where observational uncertainty due to cosmic variance is largest.
A complete analysis of cosmic variance at the faint end, therefore, requires a simulation that can resolve such low-mass halos in a box large enough to sample a wide range of local reionization histories, and treat the interplay between their star formation and the back-reaction of ionizing radiation self-consistently.
{ 
Specifically, as we will show, the faint-end HFF observations probe halos with masses $M \gtrsim 10^{8.5}~\text{M}_\odot$, so we require a simulation with a dark matter particle mass of $m_\textsc{dm} \lesssim 10^6~\text{M}_\odot$, to resolve the formation of such halos with at least a few hundred particles each.
Furthermore, since the volume searched by an HFF survey is $\sim 3000~\text{cMpc}^3$, we require a simulation with a box size of $L_\text{box} \gtrsim 100$ cMpc, so that it contains a statistically-meaningful sample of at least a few hundred survey volumes, with a self-consistent distribution of overdensities and reionization histories.
While several radiation-hydrodynamics reionization simulations have been produced in recent years \citep[e.g.][]{GF06,FDO11,Iliev14,Gnedin14,So14,Xu16,Ocvirk16,Pawlik17,Pallottini17,Aubert18,Rosdahl18,Ocvirk20,Lewis22, Kannan22}, only a few meet these size and resolution requirements, simultaneously -- namely, the Cosmic Dawn (``CoDa'') \citep{Ocvirk16,Aubert18,Ocvirk20,Lewis22} 
and \textsc{thesan} \citep{Kannan22} simulations (see Table~\ref{tab:sims}),
though, until now, these have not been analyzed for this purpose. 

\begin{table}
    \centering
    \begin{tabular}{l|c|c}
        \hline 
        \hline 
        simulation & $m_\textsc{dm}$ (M$_\odot$) & $L_\text{box}$ (cMpc) \\
        \hline 
        CoDa I     & $3.49 \times 10^5$          & 91.4 \\
        CoDa I-AMR & $2.79 \times 10^6$          & 91.4 \\
        CoDa II    & $4.07 \times 10^5$          & 94.4 \\
        CoDa III   & $5.09 \times 10^4$          & 94.4 \\
        THESAN-1   & $3.12 \times 10^6$          & 95.5 \\
        \hline
    \end{tabular}
    \caption{  A comparison of the dark matter particle masses ($m_\textsc{dm}$) and box sizes ($L_\text{box}$) of high-resolution, large-scale, radiation-hydrodynamics EOR simulations.}
    \label{tab:sims}
\end{table}

In what follows, therefore, we present the first study of the inhomogeneous UVLF during the EOR based upon a self-consistent radiation-hydrodynamics simulation of fully-coupled galaxy formation and reionization with the required large volume and high mass-resolution described above.  
We use the second-generation CoDa II simulation \citep[][]{Ocvirk20}\footnote{ 
Some results for the high-$z$ UVLF from the CoDa~II simulation were presented by us before, in \citet{Ocvirk20}, but here we will analyze its \textit{inhomogeneity} for the first time, while also presenting fitting formulae for the globally-averaged UVLF for direct comparison with the observed UVLF, with special attention to evidence for flattening at the faint end.
} to determine how the UVLF at $z\ge 6$ varies spatially in correlation with regional variations in the halo mass function and the local timing of reionization, to establish the cosmic variance of the UVLF on scales large and small in a statistically meaningful way. 
We will compare our predicted UVLF's with observations and assess the implications of our results for surveys based upon HFF lensing data.}

Our paper is organized as follows.
In \S\ref{sec:codaii}, we briefly describe the CoDa~II simulation and its relevant post-processing for this work.
In \S\ref{sec:patchy}, we illustrate the temporal evolution and spatial inhomogeneity of the UVLF in CoDa~II across a wide range of local reionization histories and overdensities.
In \S\ref{sec:HFF}, we compare our results to the HFF observations, and demonstrate a substantial discrepancy between the estimate of uncertainty due to cosmic variance derived from our simulation and that used in previous studies, when applied to the small-volume lensing results discussed here.
In \S\ref{sec:fit}, we fit our globally-averaged CoDa~II UVLF at $z=6,7,8,10,15$ to Schechter functions with and without modifications to the faint-end behavior, to determine whether our simulation predicts a flattening of the faint-end slope.
We conclude and summarize our results in \S\ref{sec:conclusion}.

\begin{table}
    \centering
    \begin{tabular}{c|c}
        \hline\hline
        parameter & value \\
        \hline
        $h$ & 0.677 \\
        $\Omega_m$ & 0.307 \\
        $\Omega_\Lambda$ & 0.693 \\
        $\Omega_b$ & 0.048 \\
        $\sigma_8$ & 0.829 \\
        $n$ & 0.963 \\
        \hline
    \end{tabular}
    \caption{Cosmological parameters from \citet{Planck14}, which are used in CoDa~II.}
    \label{tab:params}
\end{table}

\section{CoDa~II Simulation}
\label{sec:codaii}

{ 
CoDa II, described in detail in \citet{Ocvirk20}, is the second-generation radiation-hydrodynamics simulation of fully-coupled galaxy formation and reionization in a $\Lambda$CDM universe by \textit{The Cosmic Dawn (``CoDa'') Project}, based upon the massively-parallelized, hybrid CPU-GPU code \textsc{ramses-cudaton}.  CoDa II has periodic boundary conditions in a cubic volume 94.4 cMpc on a side, with $4096^3$ $N$-body particles for the dark matter and $4096^3$ grid cells for the baryonic gas and radiation field, resolving the formation of the full range of atomic-cooling halo (``ACH'') masses,  $M\gtrsim 10^8~\text{M}_\odot$, and simulating through the end of reionization to $z=5.8$.
The simulation adopts cosmological parameters from \cite{Planck14}, which are provided in Table~\ref{tab:params}.

Hydrodynamics and $N$-body dynamics are handled by the \textsc{ramses} code \citep{Teyssier02}, which uses a second-order Godunov scheme Riemann solver for the gas and a particle-mesh integrator for the dark matter.
Radiative transfer (``RT'') and thermochemistry are handled by the \textsc{aton} code \citep{AT08}, which relies on a moment-based description of the radiative transfer equations and uses the M1 closure relation \citep{GAH07}.
It tracks the out-of-equilibrium ionizations and cooling processes involving atomic hydrogen. 
Radiative quantities (energy density, flux, and pressure) are described on a fixed, comoving, Eulerian grid -- the same grid as is used for its particle-mesh $N$-body gravity solver -- and evolved according to an explicit scheme under the constraint of a CFL condition.

The latter condition is especially challenging for the cosmic reionization problem, since weak, R-type ionization fronts, driven by UV starlight emitted inside galaxies, break out of the galaxies and accelerate in the low-density IGM up to velocities that are many thousands of km/s, even approaching an appreciable fraction of the speed of light.
As a result, the time-step upper-limit set by the CFL in the presence of such high-speed I-fronts is orders of magnitude smaller than that set by the CFL condition for hydrodynamics alone (without RT), even if the latter hydro assumes optically-thin photoionization that raises the sound speed of ionized gas by heating it to $10^4$ K.   The small time steps required by the CFL when hydro and RT are fully coupled has the unfortunate consequence that the number of time-steps required to integrate over a given interval of cosmic time by finite-differencing the hydro, gravity, and RT equations together (with the same time-step) is orders of magnitude larger than for a cosmological simulaton of hydro and gravity without RT.  It is currently computationally infeasible to do this on the scale required for as large a simulation as CoDa II.  

The \textsc{ramses-cudaton} code was specifically developed to overcome this obstacle.  
It is unique in solving this problem, by being coded to run on a massively-parallel, hybrid CPU-GPU supercomputer like Titan at Oak Ridge OLCF, in which each of its thousands of nodes have, not only dozens of CPUs, but also GPUs.  
In the same wall-clock time it takes to advance the hydro and gravity equations on the CPUs for one hydro-gravity time-step, \textsc{ramses-cudaton} uses the GPUs to advance through $\sim100$ sub-steps of the RT and ionization rate equations, as well.   This enables the net  computational time of the problem with RT to approach the computational time of the problem with no RT, by speeding it up by two orders of magnitude.         

Other simulations of  reionization and galaxy formation with fully-coupled hydro and RT that solve the RT equations by a moment method, as \textsc{ramses-cudaton} does, have attempted to side-step this severe requirement of extremely small RT-step-size dictated by the CFL condition by replacing the true speed of light by an artificially reduced value -- the so-called reduced-speed-of-light approximation ``RSLA'' -- to ``trick'' the CFL into allowing larger time steps for the RT. 
This is not necessary for \textsc{ramses-cudaton}.  As a result, CoDa II was able to adopt the full speed of light, thereby avoiding the well-known artifacts introduced in the other reionization simulations by their adoption of the RSLA \citep[see][]{Deparis19,Ocvirk19}. 
Nevertheless, to simulate through the end of the EOR, down to redshift 5.8, CoDa II had to run for about 6 days on 16,384 nodes of the Titan supercomputer, using 4 cores and 1 GPU per node, for a total of 65,536 cores, with each node hosting 4 MPI processes that each managed a subvolume of $64 \times 128 \times 128$ cells.

Since the mass scale of individual stars is completely unresolved by all reionization simulations, CoDa II included, star formation is modeled by a subgrid algorithm.  In CoDa II, star particles are created in each hydro cell in which the baryon overdensity exceeds 50, with the rate of change of the stellar mass density given by}
\begin{equation}
    \Dot{\rho}_\star = \epsilon_\star \frac{\rho_\text{gas}}{t_\text{ff}}
\end{equation}
where $\rho_\text{gas}$ is the baryon density, $t_\text{ff}$ is the free-fall time, and $\epsilon_\star$ is a calibration parameter referred to as the star formation efficiency, which is set to 0.02. 

{ 
We add to our subgrid star formation algorithm a parameterized ionizing photon efficiency (IPE; the number of ionizing photons released per unit stellar baryon per unit time) into the host grid cell of each star particle.  
We define this IPE as $\xi_\textsc{ipe} \equiv f_{\rm esc,\star} \xi_{\text{ph,}\textsc{imf}}$, where $f_{\rm esc,\star}$ is the stellar-birthplace escape fraction and $\xi_{\text{ph,}\textsc{imf}}$ is the number of ionizing photons emitted per Myr per stellar baryon. 
Each stellar particle is considered to radiate for one massive star lifetime $t_\star = 10$ Myr, after which the massive stars die (triggering a supernova explosion) and the particle becomes dark in the H-ionizing UV.
We adopted an emissivity $\xi_{\text{ph,}\textsc{imf}} = 1140$ ionizing photons/Myr per stellar baryon.
This is consistent with emission by our assumed $Z=0.001$ BPASS binary stellar population model \citep{Eldridge17}, with a Kroupa initial mass function \citep{Kroupa01}, assuming no dust extinction.
We used a mono-frequency treatment of the radiation with an effective frequency of 20.28 eV.
Finally, we calibrated $f_{\rm esc,\star}$ by adjusting the value in a set of smaller-box simulations, so as to obtain a reionization redshift close to $z = 6$, which led us to adopt a value of $f_{\rm esc,\star} = 0.42$.}

\begin{table}
    \centering
    \begin{tabular}{c|c|c|c|c|c|c}
        \hline\hline
         & $z=6$ & $z=7$ & $z=8$ & $z=9$ & $z=10$ & $z=15$ \\
        \hline
        $\langle X_\textsc{hii} \rangle_\textsc{v}$ & 1 -- 1.2e-5 & 5.0e-1 & 1.7e-1 & 5.3e-2 & 1.6e-2 & 2.4e-4 \\
        \hline
    \end{tabular}
    \caption{Volume-weighted global ionized fraction in CoDa~II at given $z$.}
    \label{tab:Xion}
\end{table}

\begin{figure*}
    \centering
    \includegraphics[width=\textwidth]{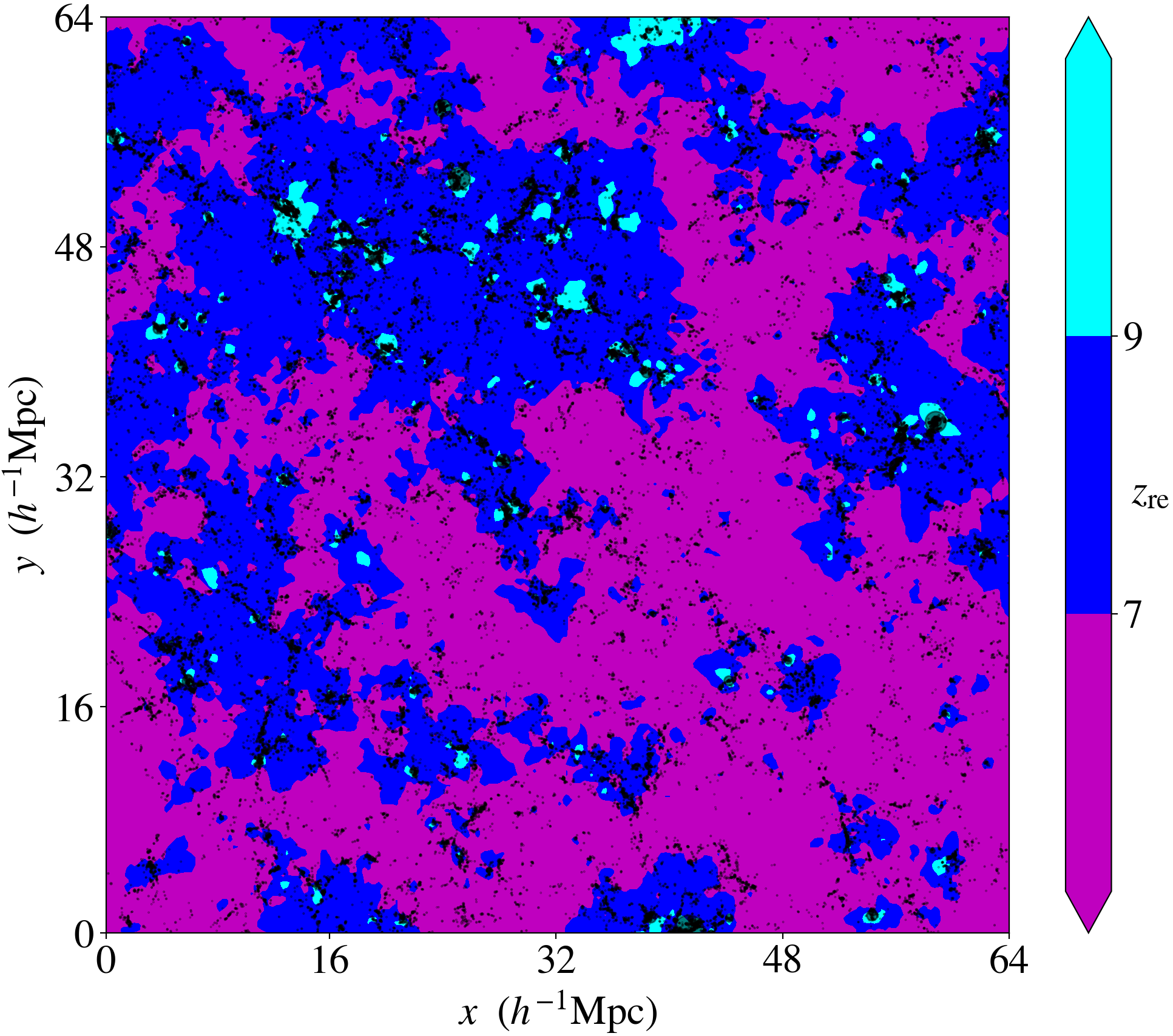}
    \caption{Contour map of a slice through the CoDa II reionization redshift field, 0.25 $h^{-1}$ cMpc thick, divided into regions that reionize relatively early ($z_\text{re} > 9$; cyan), late ($z_\text{re} < 7$; magenta), and at intermediate redshifts ($7 < z_\text{re} < 9$; blue). Black circles show the locations of halos in the same slice at $z=10$, with the size of each circle proportional to the halo's mass.}
    \label{fig:zre}
\end{figure*}

\begin{figure*}
    \centering
    \includegraphics[width=.95\textwidth]{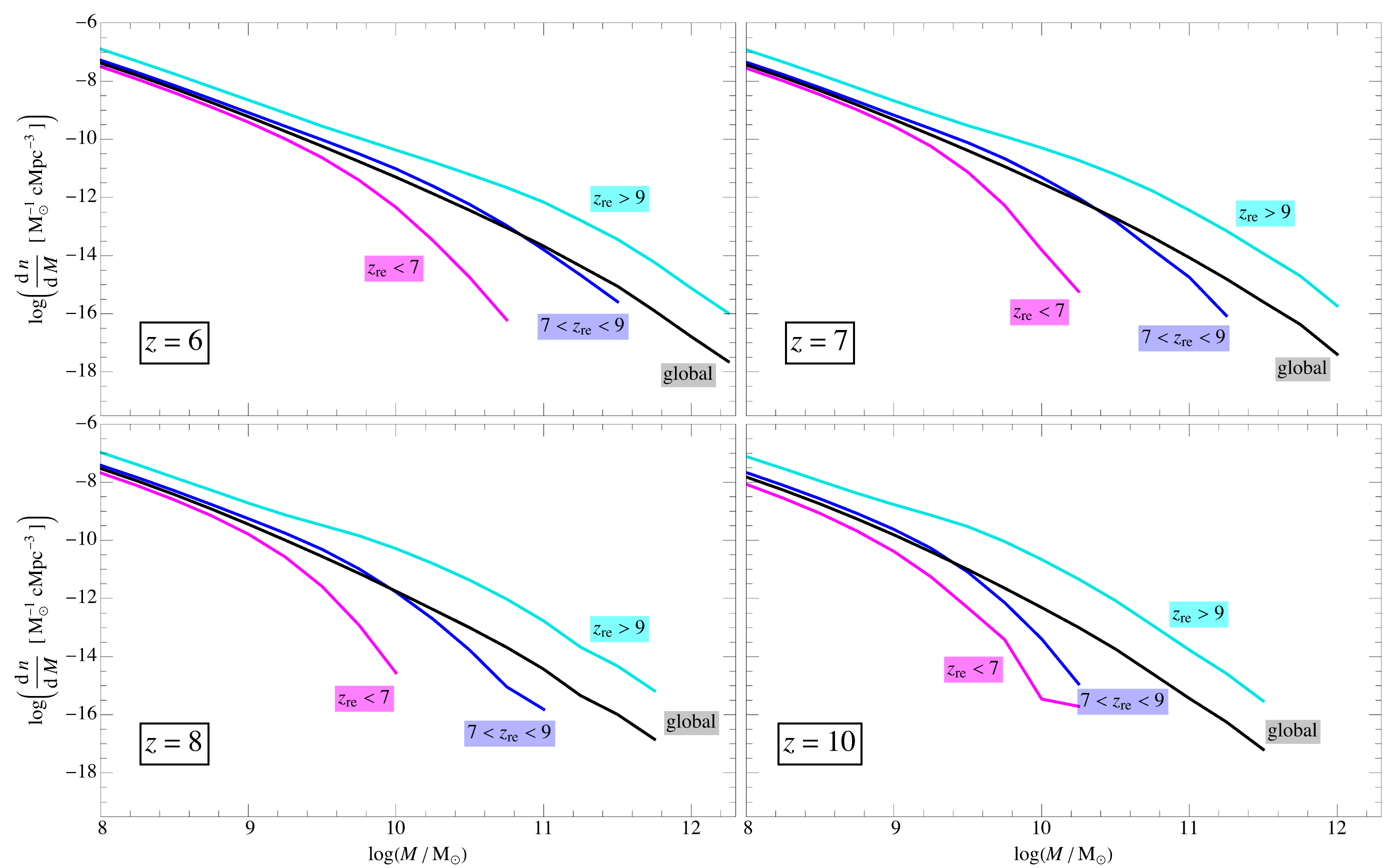}
    \caption{Halo mass functions in CoDa~II at $z=6,7,8,10$. Black solid lines show the HMF in the full simulated volume, while cyan, blue, and magenta lines show the HMF in early-, intermediate-, and late-reionizing regions, respectively, as labeled. {   
    Here and elsewhere, $\log$ implies $\log_{10}$.}}
    \label{fig:HMF}
\end{figure*}

{ 
In order compute the UVLF, we must post-process the CoDa results to find its galaxies by a dark matter halo finding algorithm, assign star particles to each host halo according to their spatial overlap with the halo volumes, and sum the emission of all the star particles associated with a given halo below the Lyman limit of H atoms to compute that galaxy's UV continuum luminosity.}
Dark matter halos are identified using a Friends-of-Friends algorithm with a standard linking length parameter of 0.2.
The mass of each halo, $M$, is defined as the total mass of all linked dark matter particles, and the virial radius is estimated as
\begin{equation}
    R_{200} = \left( \frac{3 M}{4\pi \times 200 \Bar{\rho}_\textsc{dm}} \right)^{1/3}
\end{equation}
where $\Bar{\rho}_\textsc{dm}$ is the cosmic mean dark matter density. 
Star particles are then assigned to halos if they fall within the halo's virial radius, and the masses and ages of each halo's star particles are used to compute the halo's UV luminosity and magnitude ($M_\text{UV}$) at 1600 \AA{}, according to the $Z=0.001$ BPASS binary stellar population model described above, again assuming no dust extinction.

To track and analyze the progress, patterns, and patchiness of reionization we construct the reionization redshift field of the CoDa~II simulation, illustrated in Fig.~\ref{fig:zre}.
We start by coarsening the simulated grid to $256^3$ cells, and computing the volume-weighted average ionized fraction in each of these cells at each snapshot.
(See Table~\ref{tab:Xion} for the global ionized fraction at select redshifts.)
The purpose of this coarsening is to smooth over the interiors of halos, which can be shielded from ionizing radiation due to their high densities, and instead probe the ionization state of the IGM.
Then, we identify the redshift at which each coarse-grained cell first reaches an ionized fraction of 90\%, which we define as the cell's reionization redshift, $z_\text{re}$.
Correspondingly, we identify each halo's $z_\text{re}$ as that of the coarse-grained cell its center of mass belongs to.
For the purposes of this work, we consider three ranges of reionization redshift -- $z_\text{re} > 9$, $7 < z_\text{re} < 9$, and $z_\text{re} < 7$ -- which we refer to as early-, intermediate-, and late-reionizing regions, respectively.
Fig.~\ref{fig:zre} shows a contour map of a slice through the reionization redshift field divided into these three ranges (cyan, blue, and magenta, respectively), along with the positions of $z=10$ halos (black dots) in the same slice.
There is a clear correlation between halo number density and $z_\text{re}$, with the earlier reionizing regions containing a higher density of halos than the later reionizing regions.
For example, while the early-reionizing regions are the rarest and most compact, occupying only around 2\% of the total volume, they contain around 13\% of all halos at $z=10$ (see Table~\ref{tab:frac}).
On the other hand, the vast late-reionizing regions, which occupy 58\% of the volume, contain around 32\% of the halos at $z=10$.
We explore this correlation further in the following section.

\begin{table}
    \centering
    \begin{tabular}{c|c|c|c|c|c}
        \hline\hline
         &  & $z=6$ & $z=7$ & $z=8$ & $z=10$ \\ 
        $z_\text{re}$ bin & volume & halo & halo & halo & halo \\
         & fraction & fraction & fraction & fraction & fraction \\
        \hline
        $>9$ & 0.022 & 0.074 & 0.081 & 0.089 & 0.125 \\
        7--9 & 0.395 & 0.506 & 0.496 & 0.517 & 0.556 \\
        $<7$ & 0.583 & 0.420 & 0.422 & 0.393 & 0.319 \\
        \hline
    \end{tabular}
    \caption{The fraction of the total volume occupied by early-, intermediate-, and late-reionizing regions, along with the fraction of all halos contained in these regions.}
    \label{tab:frac}
\end{table}

\begin{figure*}
    \centering
    \includegraphics[width=.95\textwidth]{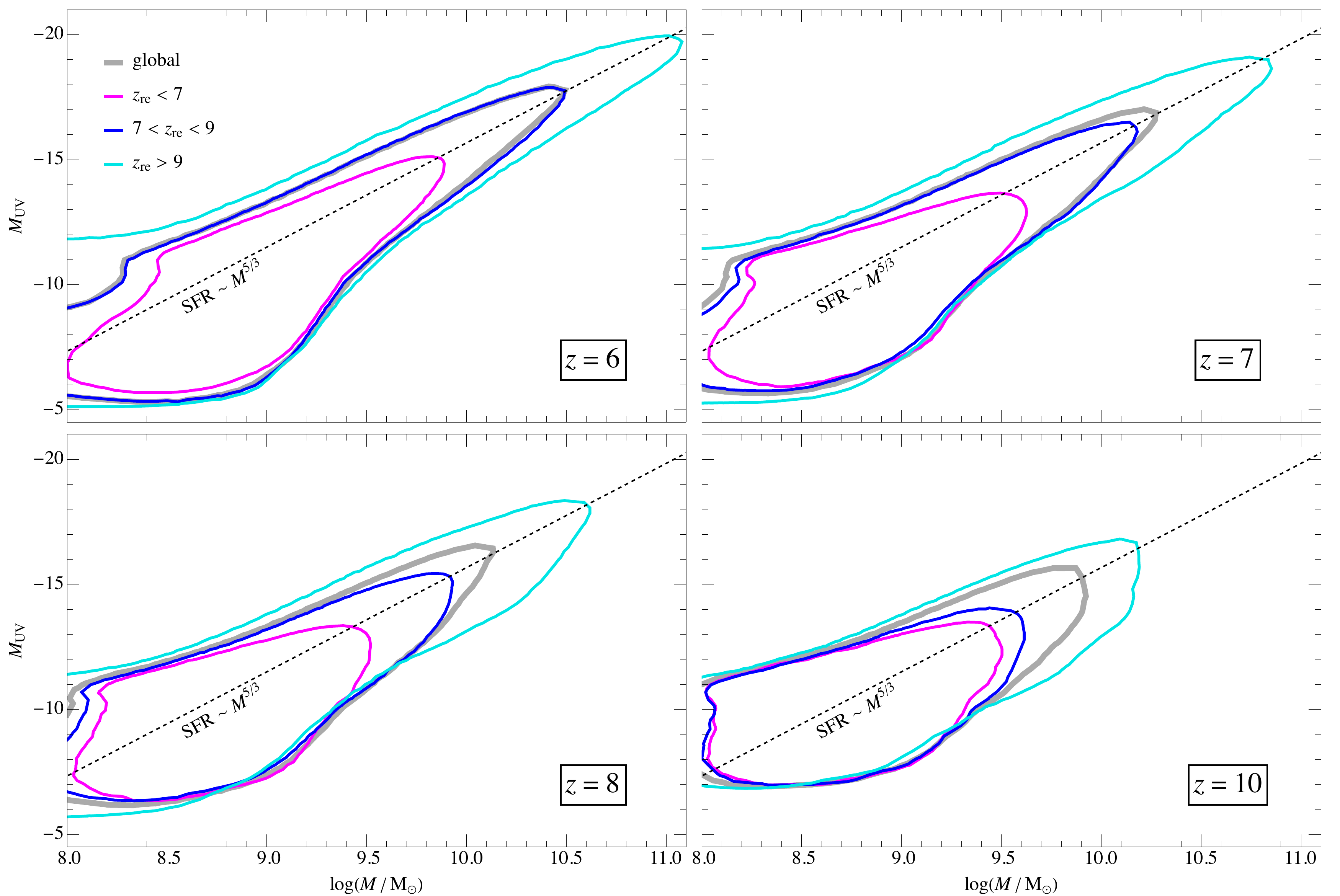}
    \caption{95\% contours for UV magnitude vs. halo mass of luminous galaxies in CoDa~II, binned by $z_\text{re}$, for the same 4 redshifts as in Fig.~\ref{fig:HMF} and using the same color lines for different $z_\text{re}$ bins. The thick grey line is the global median. The black dashed line is the rough expectation given a (pre-suppression) SFR~$\propto M^{5/3}$.}
    \label{fig:mass-mag}
\end{figure*}

\section{CoDa~II UV Luminosity Function}

\subsection{The Imprint of Patchy Reionization}
\label{sec:patchy}

In general, the UVLF can be decomposed into two factors: the halo mass function, and the star formation rate of halos as a function of mass.
As we described in \cite{Dawoodbhoy18}, both of these factors are strongly correlated with the reionization history of the region in which they are observed, and so too must the UVLF be.

The earliest regions to reionize will be those that are the most dense, since these regions will be the first to form a large number of star-forming galaxies -- the primary sources of reionization.
The latest regions to reionize will be the voids, which typically do not form enough stars to reionize themselves, and so require `importing' ionizing radiation from external, earlier-reionizing regions nearby, in order for them to become reionized.
Therefore, there is a positive correlation between the reionization redshift of a region and its halo mass function (HMF; the number density of halos per unit mass): higher-$z_\text{re}$ regions have higher HMFs that extend out to higher mass. 
We show the combined HMFs in regions binned by their $z_\text{re}$, for four different redshifts, in Fig.~\ref{fig:HMF}.
As can be seen, the early-reionizing regions ($z_\text{re} > 9$) have the highest and most extended (i.e. the turn-over to a steeper decline occurs at a higher mass) HMF at all redshifts, followed by the intermediate- ($7 < z_\text{re} < 9$) and late-reionizing ($z_\text{re} < 7$) regions.
The globally averaged HMF tracks closest to the intermediate-reionizing regions at the low-mass end.

Naturally, a higher HMF will result in a higher UVLF, overall, so we should expect the correlation between HMF and $z_\text{re}$ to translate to a correlation between UVLF and $z_\text{re}$.
However, the actual UV luminosity of each halo is determined by its star formation rate (SFR), which has a more complicated relationship with $z_\text{re}$. 
First, for relatively high-mass halos ($M\gtrsim 10^9~\rm{M}_\odot$), the SFR scales as $\sim M^{5/3}$ \citep[see, e.g.,][]{Ocvirk16,Ocvirk20}, which means HMFs that are more extended to high mass (i.e. those of earlier reionizing regions) will correspond to UVLFs that are more extended to the bright end (i.e. the turn-over to a steeper decline occurs at a brighter magnitude).
We illustrate the effect of this SFR scaling in Fig.~\ref{fig:mass-mag}, which shows 95\% contours for the UV magnitude vs. halo mass of luminous galaxies in CoDa~II, binned by $z_\text{re}$, at $z = 6-10$.
The rough expectation from the SFR $\sim M^{5/3}$ scaling (i.e. assuming UV luminosity is proportional to SFR) is well-obeyed for $M\gtrsim 10^{9.5}~\rm{M}_\odot$.
Furthermore, in addition to the earlier reionizing regions having more halos at all masses and a HMF that extends out to higher mass, there is also a higher fraction of halos of a \textit{given} mass at a given redshift with brighter UV magnitudes in earlier reionizing regions than in later reionizing regions (e.g. notice that the bright edge of the contours are `stacked' by $z_\text{re}$), which further contributes to the difference in their UVLFs.

\begin{figure}
    \centering
    \includegraphics[width=\columnwidth]{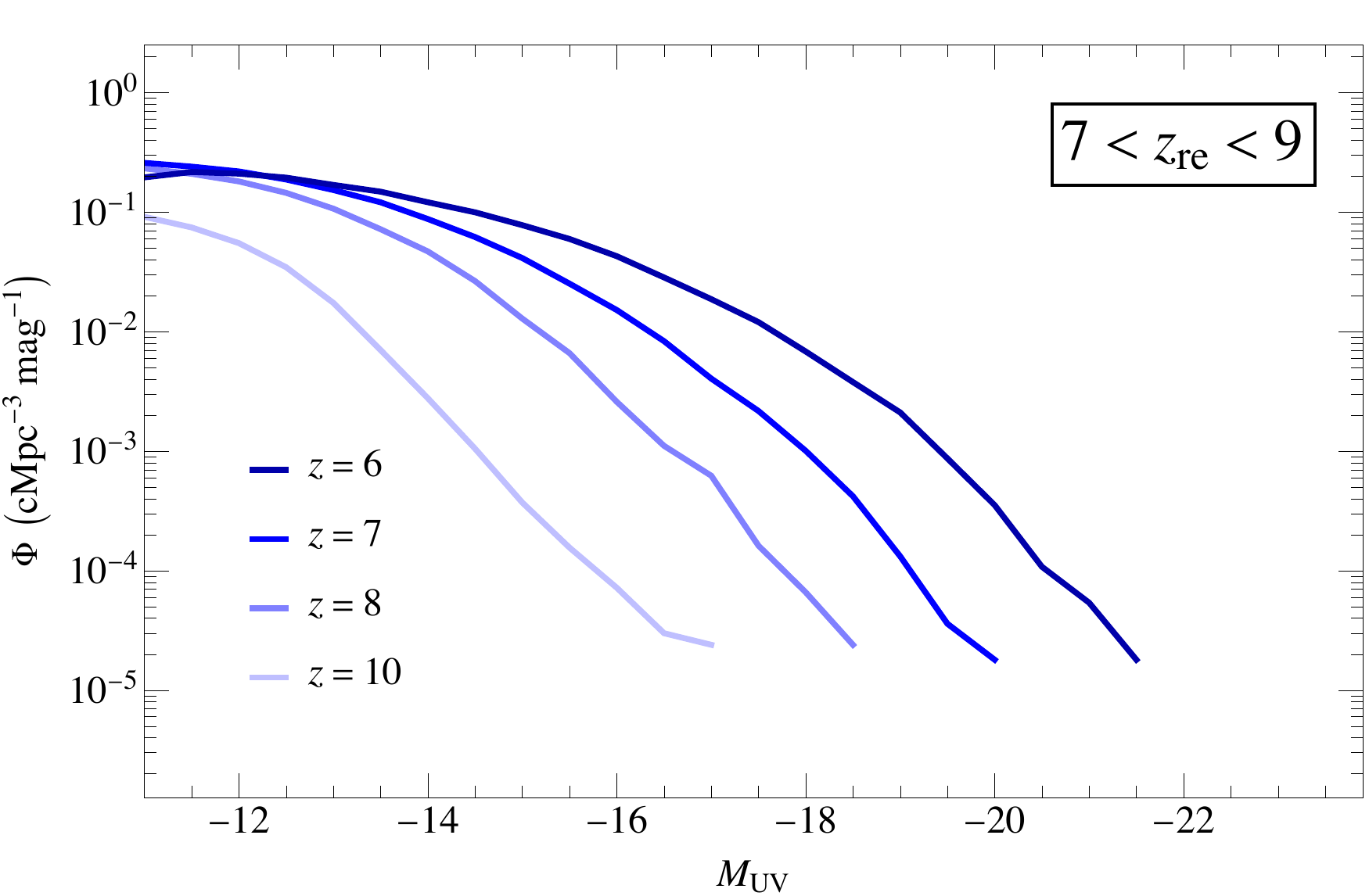}
    \caption{The evolution of the UVLF in intermediate-reionizing regions $(7 < z_\text{re} < 9)$, from pre-reionization ($z=10$) to post-reionization ($z=6$). Notice that the latter features a much flatter faint-end than the former.}
    \label{fig:UVLF-midZre}
\end{figure}

\begin{figure*}
    \centering
    \includegraphics[width=\textwidth]{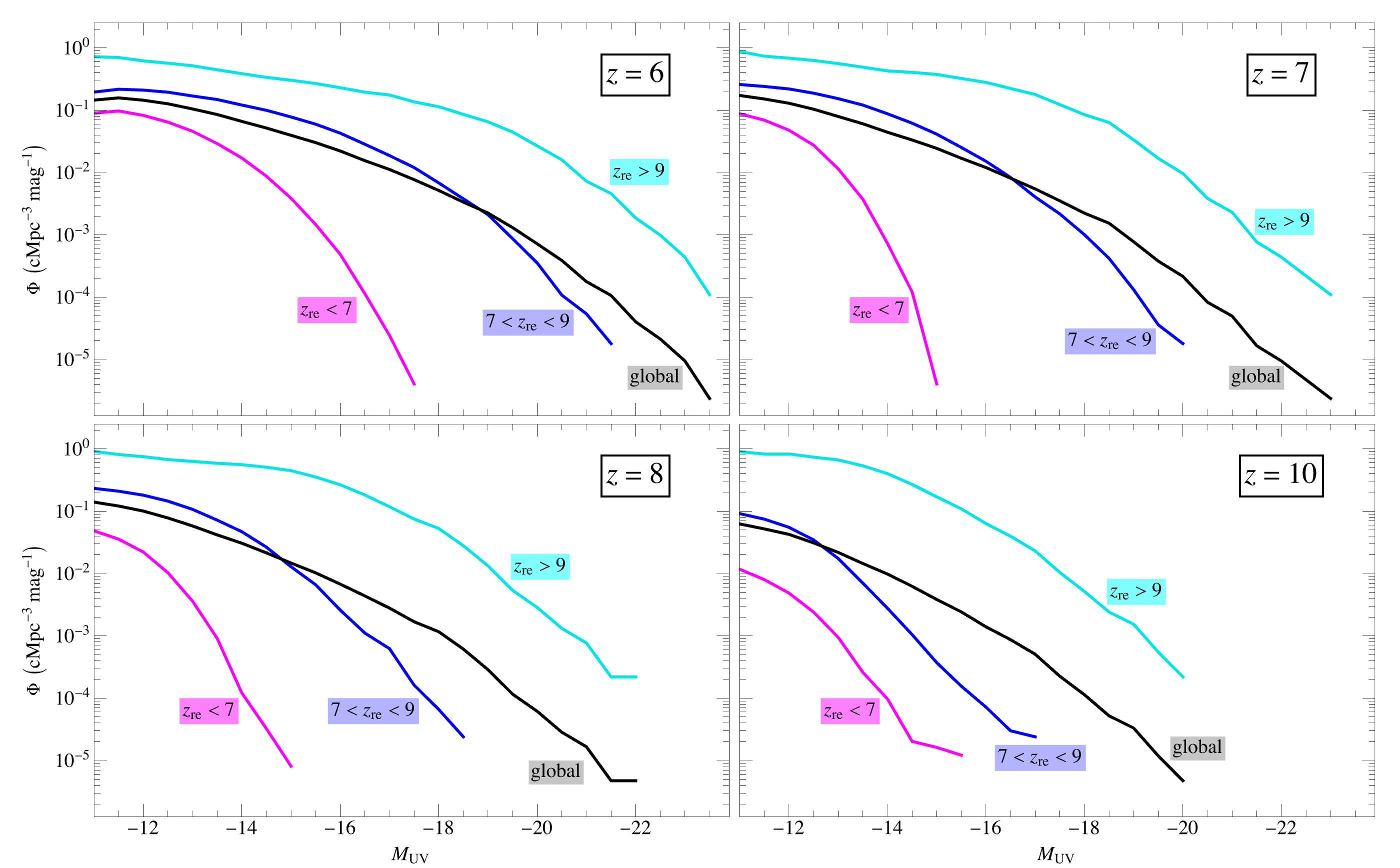}\\
    \caption{UV luminosity functions in CoDa~II at $z=6,7,8,10$. Black lines represent the full simulated volume, while cyan, blue, and magenta lines show early-, intermediate-, and late-reionizing regions, respectively. 
    }
    \label{fig:UVLF}
\end{figure*}

On the other hand, lower mass halos ($M \lesssim 10^{9.5}~\rm{M}_\odot$)\footnote{Note that the flattening seen at $M \lesssim 10^{8.5}~\rm{M}_\odot$ in Fig.~\ref{fig:mass-mag} is likely a resolution-limit effect.}
deviate from the $M^{5/3}$ scaling at low redshift, due to the suppression of star formation in low-mass halos, caused by reionization feedback (e.g. notice that the faint edge of the contours drop more sharply than the $M^{5/3}$ scaling for $M \lesssim 10^{9.5}~\rm{M}_\odot$ at late redshift).
After a region becomes reionized, low-mass halos are unable to accrete the photo-ionized gas in the IGM, due to its increased temperature, and so they will no longer have the fuel required to form stars \citep[see, e.g.,][]{Dawoodbhoy18,Ocvirk20}.
Since these halos populate the faint-end of the UVLF prior to their local reionization, we should expect to see a reduction at this faint-end over time as reionization occurs and the suppressed halos move to fainter magnitudes (or disappear entirely), which is usually characterized in terms of a ``turn-over'' in the faint-end slope.
{ For example, in their semi-analytical study of inhomogeneous reionization feedback, \citet{KC11} found such a turn-over for $M_\text{UV} \gtrsim -17$ at $z=8$, preferentially in the UVLFs of overdense regions (which reionize relatively early).}
To illustrate this effect in our simulation, we show the UVLF of intermediate-reionizing regions over time in Fig.~\ref{fig:UVLF-midZre}.
Notice the change in slope and curvature over time at magnitudes $-16 \lesssim M_\text{UV} \lesssim -11$.
Prior to these regions' local reionization (i.e $z=10$), the faint-end slope is fairly steep, roughly following $\Phi \propto M_\text{UV}^{0.4}$ for $-13 \lesssim M_\text{UV} \lesssim -11$. During local reionization, however, the faint-end slope gradually flattens out, and by the time reionization has ended for these regions (i.e. $z=6$), the faint-end power law index is close to 0 in this magnitude range.

Consequently, the UVLF one observes depends on \textit{where} one looks -- an early-reionizing patch of the Universe will have a relatively high and bright-end-extended UVLF, whereas a late-reionizing patch will have a relatively low and bright-end-compressed UVLF -- and also \textit{when} one looks -- a region that is observed prior to its local reionization will have a relatively steep faint-end slope, whereas a region observed after its local reionization will have a relatively flat faint-end slope.
We show these trends in Fig.~\ref{fig:UVLF}, which plots the CoDa~II UVLFs for early-, intermediate-, and late-reionizing regions at four redshifts, along with the global average.

An important implication of these results is that small-volume observations of the UVLF will necessarily be biased in one way or another, due to the strong correlations with local density and reionization redshift.
For example, observations that search in uniformly random volumes are likely to be probing voids, which are underdense regions, since they  
occupy the most volume.
As a result, such observations are likely to return UVLFs that are lower and less extended at the bright end than the cosmic mean.
Furthermore, since these regions reionize relatively late, the inferred UVLFs are likely to have steeper-than-average faint-end slopes.
On the other hand, observations that preferentially search near the brightest sources are likely to be probing highly overdense regions that reionize relatively early, since these regions have UVLFs that are the most extended to the bright-end.
Thus, such observations are likely to return UVLFs that are higher and more extended at the bright end than the cosmic mean, with a flatter-than-average faint-end slope.

\subsection{Implications for Faint-End HFF Observations}
\label{sec:HFF}

The analysis of the previous section illustrates the dramatic variability of the UVLF among regions with different reionization histories.
However, the volume of the $z_\text{re}$-binned cells in which the UVLF is computed is rather small, only $(250/h~\rm{ckpc})^3$. 
For many observational purposes, it is more useful to assess the variance in the UVLF on larger scales, e.g. characteristic of the size of a galaxy survey.
To that end, we divided our CoDa~II box into 256 non-overlapping subvolumes, each spanning around 3300 cMpc$^3$, which is of order the { survey volumes searched by each of the HFF lensing-cluster fields}, which have been used previously to measure the faint-end of the high-$z$ UVLF.
Each subvolume contains $32\times32\times64=65536$ of the $256^3$ coarse-grained cells used in the reionization redshift field of the previous section, and so will encompass a range of reionization histories.
{  To characterize their typical reionization history, we compute the mean reionization redshift of each subvolume in two ways: (1) a halo-weighted average $\big(\langle z_\text{re} \rangle_\text{h}\big)$, obtained by averaging over the reionization redshifts of all halos in the subvolume, and (2)~a volume-weighted average $\big(\langle z_\text{re} \rangle_\text{v}\big)$, obtained by averaging over the reionization redshifts of all coarse-grained cells in the subvolume.
The distributions of these two means across our subvolumes is shown in the top panel of Fig.~\ref{fig:zre-hist}.  

\begin{figure}
    \centering
    \includegraphics[width=\columnwidth]{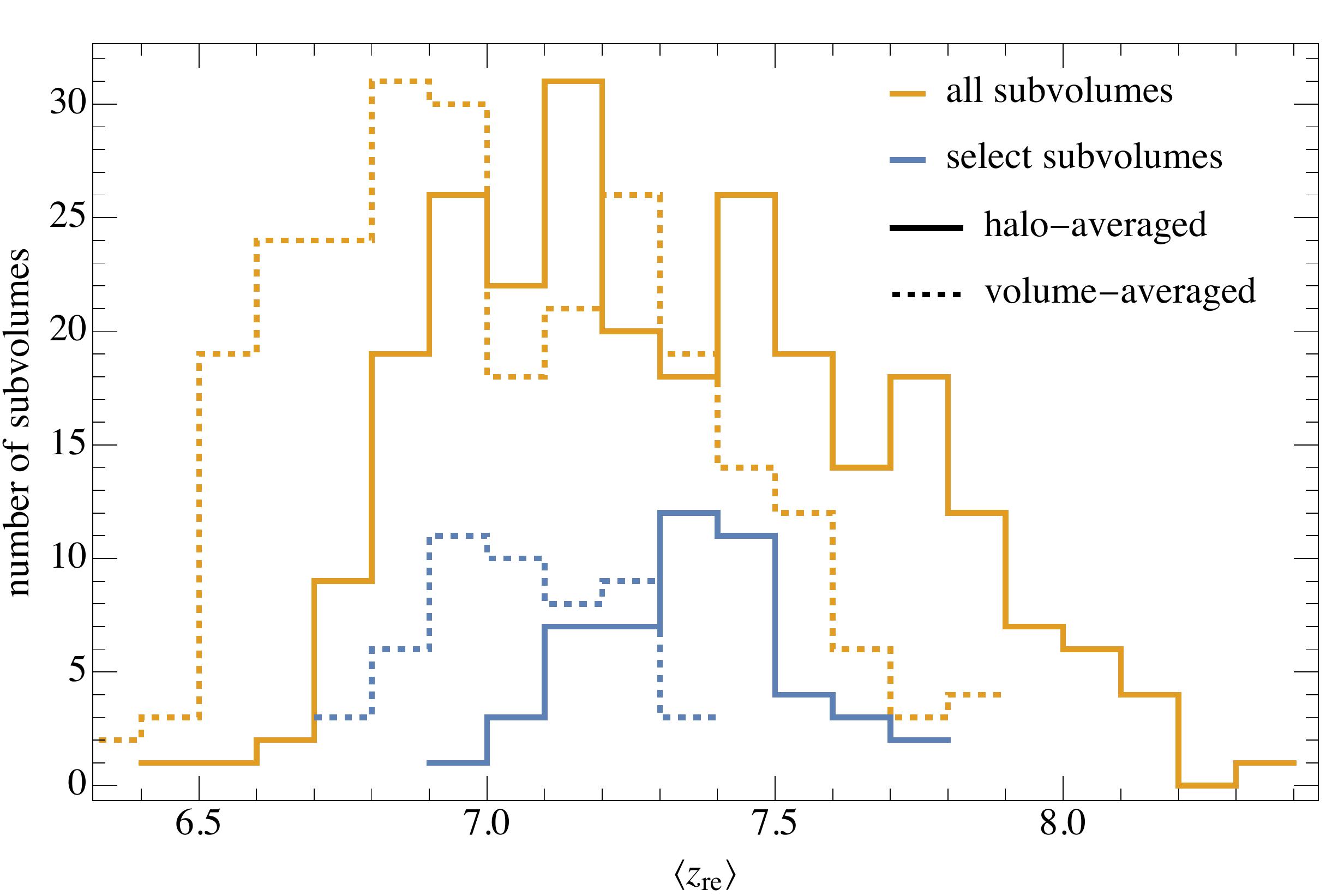}\\
    \includegraphics[width=\columnwidth]{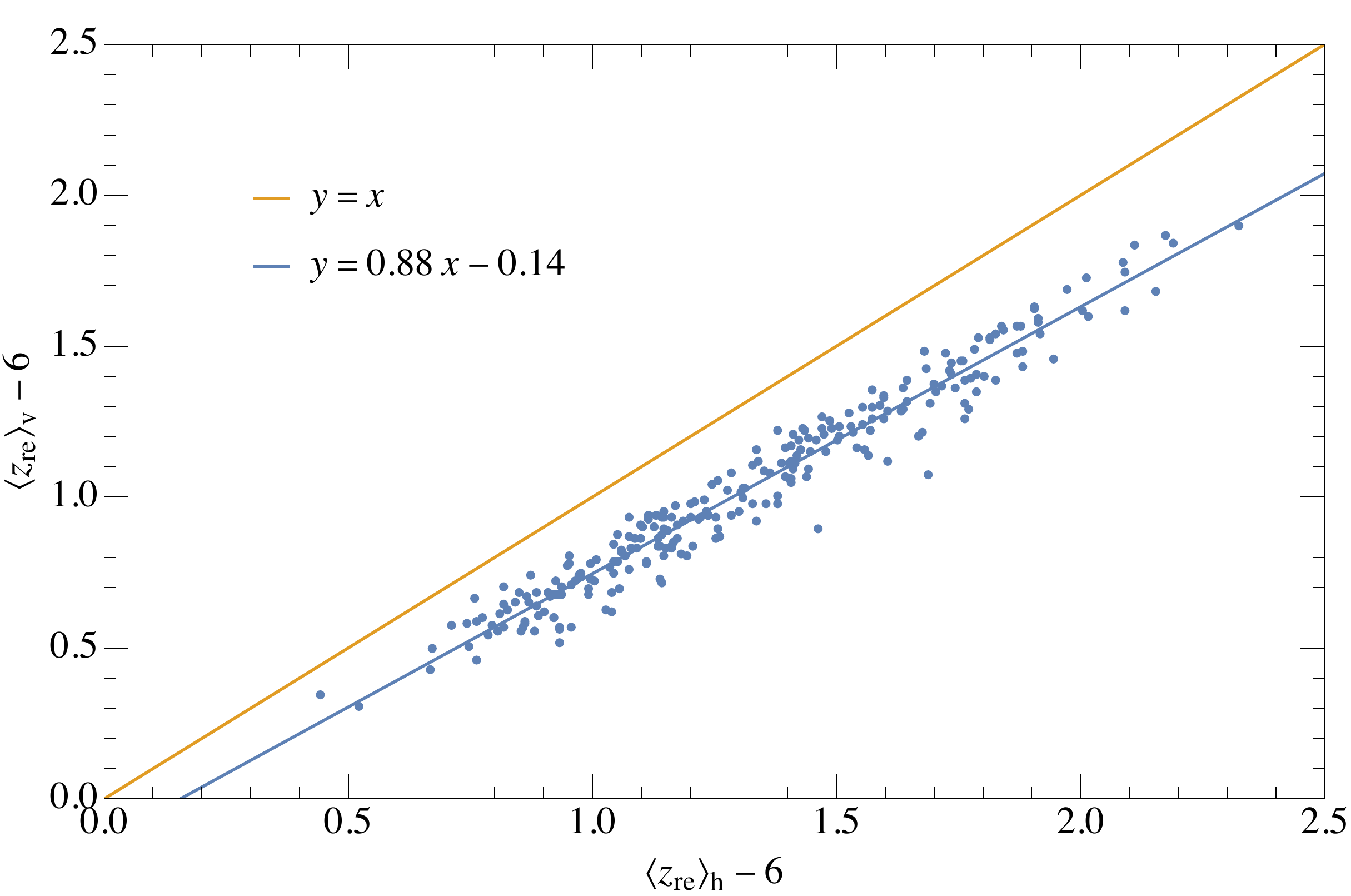}
    \caption{  Top: Histograms of the mean reionization redshifts of CoDa~II subvolumes (3300 cMpc$^3$ in size, each).
    Solid lines represent the halo-weighted average of $z_\text{re}$ across each subvolume (i.e. the average $z_\text{re}$ of all halos in each subvolume), while dashed lines represent the volume-weighted average.
    Yellow lines show the distribution for all subvolumes, while blue lines show that of a subset of 50 subvolumes that most closely match the bright-end data from \citet{Livermore17} (see Fig.~\ref{fig:select-subUVLF} and accompanying text for a description of how these 50 subvolumes are selected).
    Bottom: Volume-weighted vs. halo-weighted averages of $z_\text{re}$ across each subvolume (blue points) fit to a linear relation (blue line). All points fall below the equality line (yellow), meaning $\langle z_\text{re} \rangle_\text{v}$ is always less (i.e. later) than $\langle z_\text{re} \rangle_\text{h}$.}
    \label{fig:zre-hist}
\end{figure}

\begin{figure}
    \centering
    \includegraphics[width=\columnwidth]{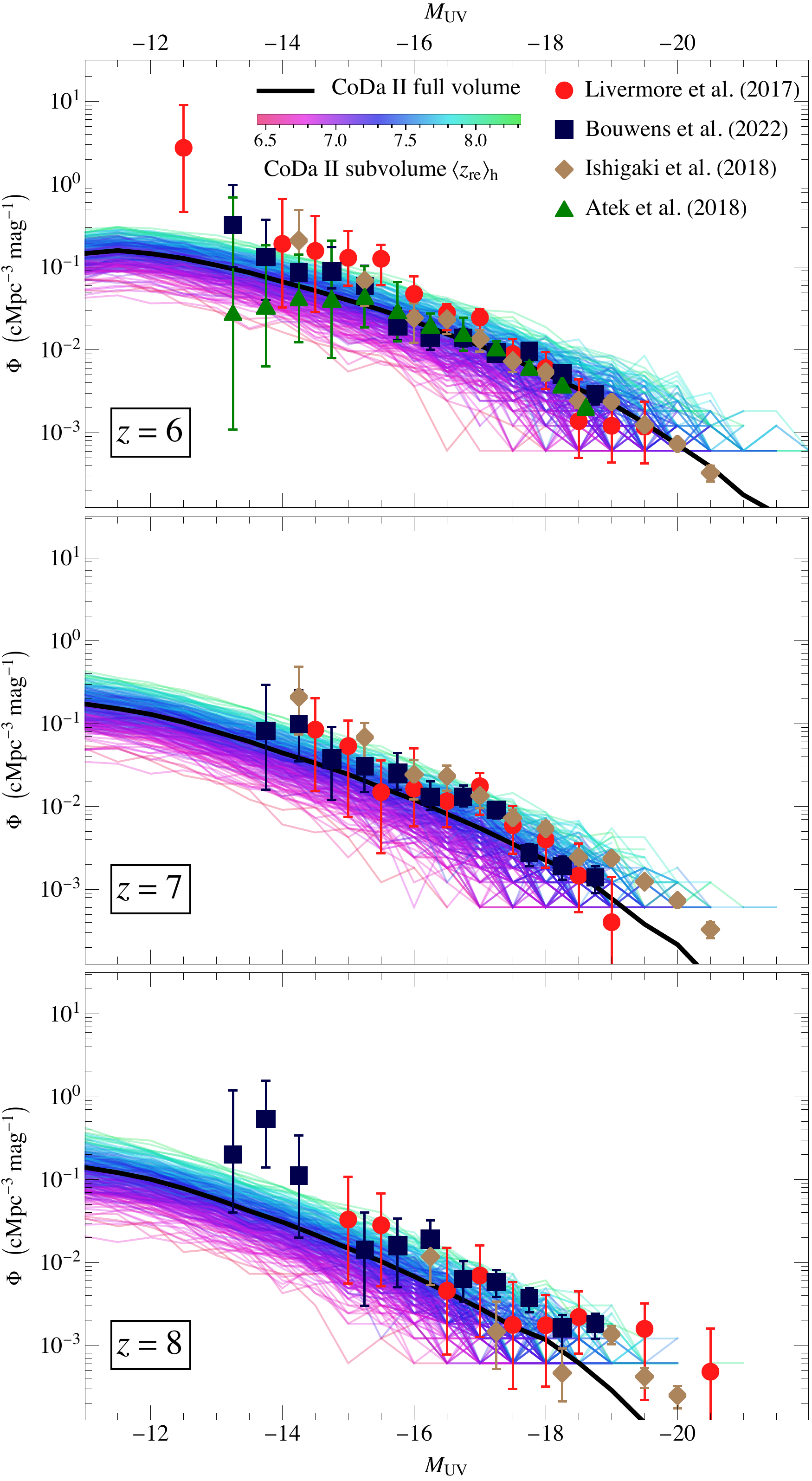}
    \caption{CoDa~II UVLFs at $z=6,7,8$ compared to observational HFF data from \citet{Livermore17,Bouwens22,Ishigaki18,Atek18}. In addition to the full volume UVLF (thick black line), we also show the UVLFs of 256 CoDa~II subvolumes -- each spanning around 3300 cMpc$^3$, which is of order the { survey volumes searched by each of the HFF lensing-cluster fields} -- as thin lines colored according to the mean reionization redshift of their constituent halos, as indicated in the legend in the top panel. The spread in the UVLFs of this collection of subvolumes is a measure of the cosmic variance on this scale.}
    \label{fig:subUVLF}
\end{figure}

As the histograms show, the distribution of volume-averaged reionization redshifts is skewed towards later redshifts than those of the halo-weighted averages. This is made clear by the plot of $\langle z_\text{re}\rangle_\text{v}$ vs. $\langle z_\text{re}\rangle_\text{h}$ for each subvolume, in the bottom panel of Fig.~\ref{fig:zre-hist}.   This trend is to be expected, since the effects of reionization tend to propagate ``inside-out'', from the neighborhoods of clustered galaxies to the surrounding, larger volumes of the IGM.  Nevertheless, Fig.~\ref{fig:zre-hist} makes it clear that, even after averaging over the full range of local reionization redshifts within a given survey volume, those survey volumes are small enough that there is still a large variation in this average reionization redshift from one survey volume to another. 
This means we should expect there to be a corresponding scatter amongst the UVLF's derived for different survey volumes of this size.} 

We plot the $z=6,7,8$ UVLFs of our subvolumes in Fig.~\ref{fig:subUVLF}.
The curves for each subvolume are colored according to their $\langle z_\text{re} \rangle_\text{h}$, with cyan corresponding to earlier-reionizing regions, magenta corresponding to later-reionizing regions, and blue in between.
Naturally, the earlier-reionizing regions have higher UVLFs, and so on.

For comparison, we also plot the HFF observational results from \cite{Livermore17} (red circles), \cite{Bouwens22} (blue squares), \cite{Ishigaki18} (brown diamonds), and \cite{Atek18} (green triangles).\footnote{
{ Note that \citet{Bouwens22}, \citet{Ishigaki18}, and \citet{Atek18} each analyze the full set of six HFF clusters, while \citet{Livermore17} analyze a subset of two of them.}
}
Their inferred UVLFs are broadly consistent with our CoDa~II subvolumes, but there is one noteworthy discrepant data point from \cite{Livermore17} at the faint-end at $z=6$.\footnote{
{ While the second-faintest point from the \citet{Bouwens22} results at $z = 8$ is also somewhat discrepant with our results, since both data points on either side of it are not, we consider this to be anomalous and do not discuss it further here.}
}
By this late redshift, most of the volume in our CoDa~II simulation has been reionized, so most low-mass halos in the subvolumes have been suppressed, and hence the faint-end slopes of the subvolumes' UVLFs are rather flat. 
However, due to the fact that they identified a single $M_\text{UV} = -12.5$ galaxy at $z=6$, \cite{Livermore17} inferred a faint-end slope that remains steep down to this low magnitude.
Galaxies that faint must be located very close to caustics in the lensing field, in order to be sufficiently magnified so as to be visible, so the effective volume searched for such a galaxy is much smaller than the full volume probed by the entire survey field. 
According to the lensing models used by \cite{Livermore17}, the effective volume searched for a $M_\text{UV} = -12.5$ galaxy is around 0.73~cMpc$^3$.
Therefore, identifying even a single galaxy in a randomly sampled volume that small implies a high UVLF at that galaxy's magnitude -- high enough to be inconsistent with the average UVLF at that magnitude in all of our CoDa~II subvolumes.
{ We note that this data point is similarly discrepant with the UVLF predicted by \textsc{thesan} \citep{Kannan22}, another large-scale, high-resolution radiation-hydrodynamics simulation that is otherwise broadly consistent with high-$z$ UVLF observations (as CoDa~II is), though they do not explore the inhomogeneity of their UVLF.}

\begin{figure}
    \centering
    \includegraphics[width=\columnwidth]{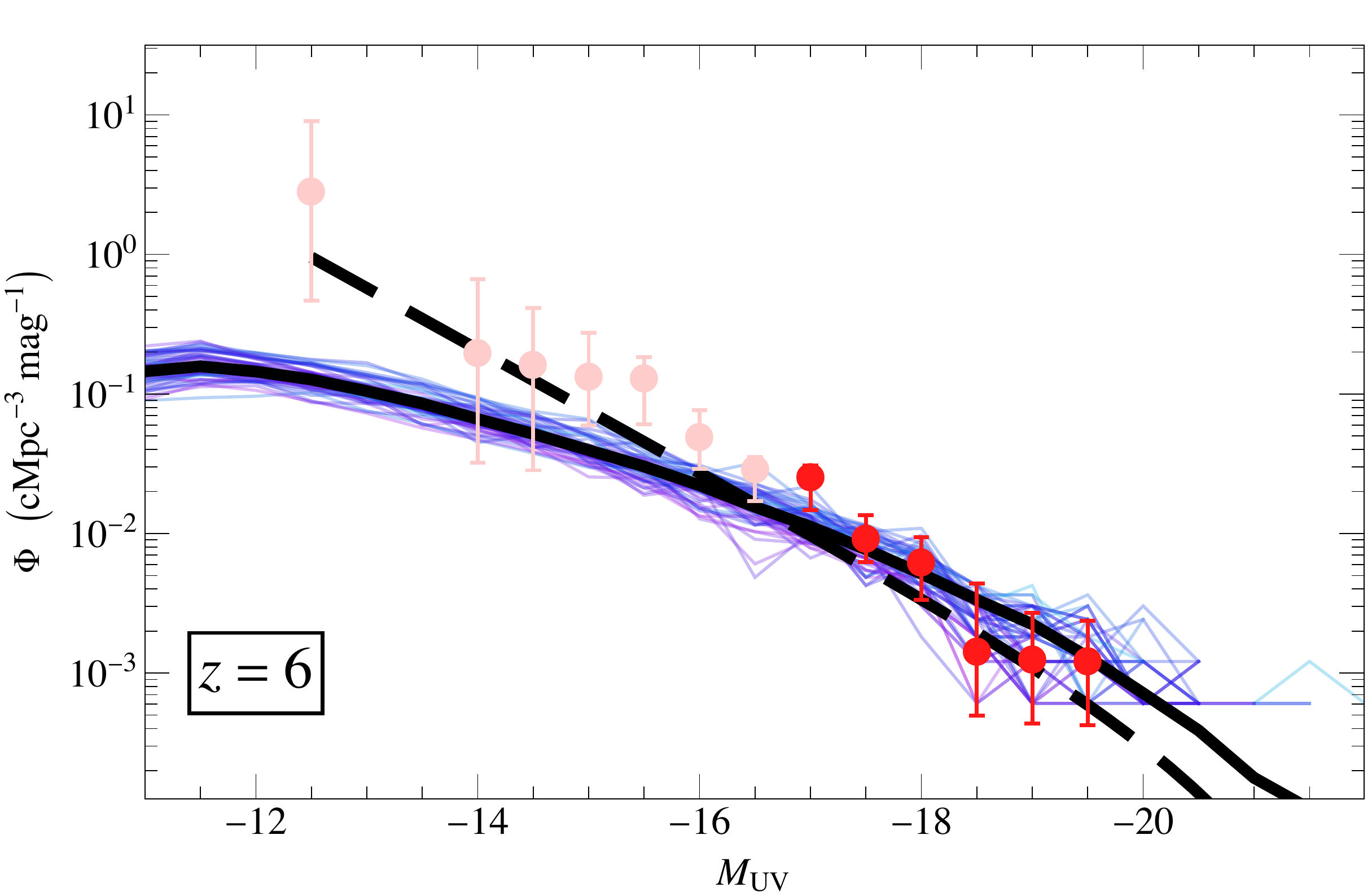}
    \caption{A comparison of the $z=6$ UVLF data from \citet{Livermore17} to the UVLFs in a selection of 50 CoDa~II subvolumes -- a subset of the 256 subvolumes shown in Fig.~\ref{fig:subUVLF}, chosen to most closely match the bright-end of the observational data (saturated red points). Despite their matching at the bright-end, the subvolumes are still discrepant with the data point at $M_\text{UV} = -12.5$. For reference, the full volume CoDa~II UVLF (thick solid line) and best-fit Schechter function from \citet{Livermore17} (thick dashed line) are also shown.}
    \label{fig:select-subUVLF}
\end{figure}

\begin{figure}
    \centering
    \includegraphics[width=\columnwidth]{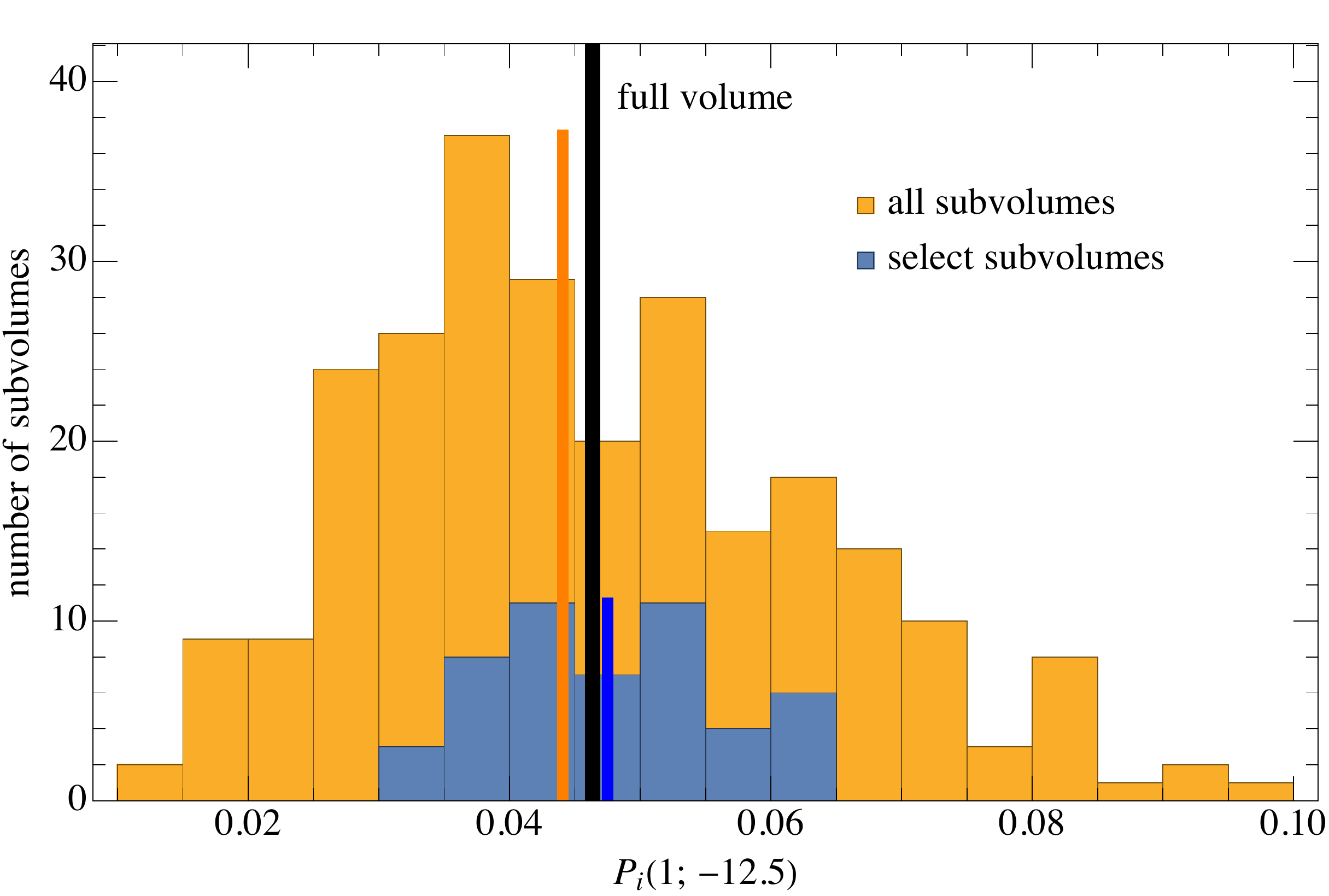}
    \caption{
    The number of subvolumes in which one $M_\text{UV} = -12.5$ galaxy may be found in a random 0.73 cMpc$^3$ region with probability $P_i(1; -12.5)$.
    Orange bars represent all 256 subvolumes in CoDa~II, while blue bars represent only those select subvolumes that most closely match the bright-end HFF data from \citet{Livermore17}. The orange and blue lines mark the median probability for their color-coordinated samples, while the black line marks the probability for the globally-averaged CoDa~II UVLF.}
    \label{fig:hist}
\end{figure}

{ To assess the degree of discrepancy between our CoDa~II results and the} observations of \cite{Livermore17}, we can estimate the probability of observing a single $M_\text{UV} = -12.5$ galaxy in a randomly sampled 0.73 cMpc$^3$ region within each CoDa~II subvolume.
For this purpose, we identified a subset of 50 of our CoDa~II subvolumes, chosen because their UVLF at brighter magnitudes most closely matches the UVLF of \cite{Livermore17} at those magnitudes, for which the effective volume surveyed matches the subvolume size, as shown in Fig.~\ref{fig:select-subUVLF}.
In particular, we selected the subvolumes with the 50 lowest $\chi^2$ values when compared to the number of observed galaxies at each magnitude in the range $-19.5 \leq M_\text{UV} \leq -17$.
The data points at these magnitudes are indicated by the saturated red circles in the figure.
{ 
Then, given the number density of $M_\text{UV} = -12.5$ galaxies in the $i^\text{th}$ subvolume,
\begin{equation}
    n_i(-12.5) \simeq \Phi_i(-12.5) \, \Delta M_\text{UV}
\end{equation} 
where $\Delta M_\text{UV} = 0.5$ \citep[as was used by][]{Livermore17},
the probability of finding $N$ such galaxies in a random search of a $V_\text{eff}(-12.5) = 0.73~\rm{cMpc}^3$ volume is given by the Poisson distribution}\footnote{
Note that there are at least 143 such galaxies in each of the select $\sim~3300$~cMpc$^3$ subvolumes, and around 209 on average. 
When considering all subvolumes, not just those selected to most closely match the bright-end of \cite{Livermore17}, there are at least 58 such galaxies in each, and around 215 on average.}
{ 
\begin{equation}
    P_i(N; -12.5) = \frac{\big(n_i(-12.5)\, V_\text{eff}(-12.5)\big)^N e^{-n_i(-12.5)\, V_\text{eff}(-12.5)}}{N!} 
\end{equation}
Thus, the probability of finding one such galaxy in the limit $n_i(-12.5) \, V_\text{eff}(-12.5) \ll 1$ is
\begin{equation}
    P_i(1; -12.5) \approx n_i(-12.5)\, V_\text{eff}(-12.5)
\end{equation}}
We find that across all of the selected subvolumes, this probability falls in the range [3.2\%, 6.5\%], with a median probability of around 4.8\%.
On the other hand, if we compute the probability across all subvolumes, not just the selected ones, we find a range of [1.3\%, 9.5\%], with a median around 4.4\%.
Using the globally-averaged CoDa~II UVLF, instead, we find a probability of 4.6\%.
We show a histogram of probability estimates for detecting a $M_\text{UV} = -12.5$ galaxy in our subvolumes in Fig.~\ref{fig:hist}.

\subsubsection{{ Cosmic Variance in Faint-End Lensing Surveys}} \label{sec:cv}

It is worth noting that there is significant uncertainty in the analysis of these extremely high-magnification faint galaxies, due to uncertainties in both the lensing models and cosmic variance, since the effective volumes searched are so small.
With regards to the former, the analysis of \citet{Bouwens17}, for example, re-interprets the $M_\text{UV} = -12.5$ galaxy as a brighter galaxy ($M_\text{UV}=-14.25$) in a larger effective search volume, with a greater uncertainty attributed to the lensing models.
With regards to the latter, \citet{Livermore17} account for uncertainty due to cosmic variance when fitting their data to Schechter functions by following the work of \citet{Robertson14}, which expresses the uncertainty in terms of the galaxy clustering bias, e.g.
\begin{equation}
    \sigma_\textsc{cv,l}(V) = \frac{ \sqrt{ \langle (n_\textsc{l}(\bm{x},V)-\Bar{n}_\textsc{l})^2\rangle_{\bm{x}} } }{\Bar{n}_\textsc{l}} = b_\textsc{l}(V) \, \sigma_\textsc{dm}(V)
\end{equation}
where $\sigma_\textsc{cv,l}(V)$ is the fractional uncertainty due to cosmic variance for galaxies of luminosity (or magnitude) $L$ observed in a volume $V$, $n_\textsc{l}(\bm{x},V)$ is the local number density of galaxies of the same luminosity in a volume $V$ located at $\bm{x}$, $\Bar{n}_\textsc{l}$ is the global mean number density of galaxies of the same luminosity, $\langle\quad\rangle_{\bm{x}}$ denotes a global average over all locations $\bm{x}$, $b_\textsc{l}(V)$ is the bias of galaxies of the same luminosity on the scale of volume $V$, and $\sigma_\textsc{dm}(V)$ is the linearly-extrapolated RMS of dark matter density fluctuations on the same scale.
\citet{Robertson14} use, as their estimate of the bias, the analysis of dark matter simulations in \citet{Tinker10}. 
However, since the focus of \citet{Tinker10} was on \textit{large-scale} bias -- i.e. in the limit where the bias is scale-independent, $b_\textsc{l}(V) \to b_\textsc{l}$ -- their analysis was restricted to only the 5--10 largest wavelength modes in each simulation, with simulation box lengths in the range 80--1280~$h^{-1}$~cMpc.
Therefore, we believe their results will underestimate the bias on the \textit{very small scales} probed by the extremely high-magnification regions of the HFF data, for which the bias is scale-dependent. 
{  As a consequence, the use of this bias in \citet{Robertson14}
will underestimate the uncertainty due to cosmic variance on such small scales, as well.  Since detection of the faintest galaxies requires the greatest magnification, the use of observations like the \citet{Livermore17} $M_\text{UV} = -12.5$ galaxy to infer the UVLF is especially affected by this underestimation.}

\begin{figure}
    \centering
    \includegraphics[width=\columnwidth]{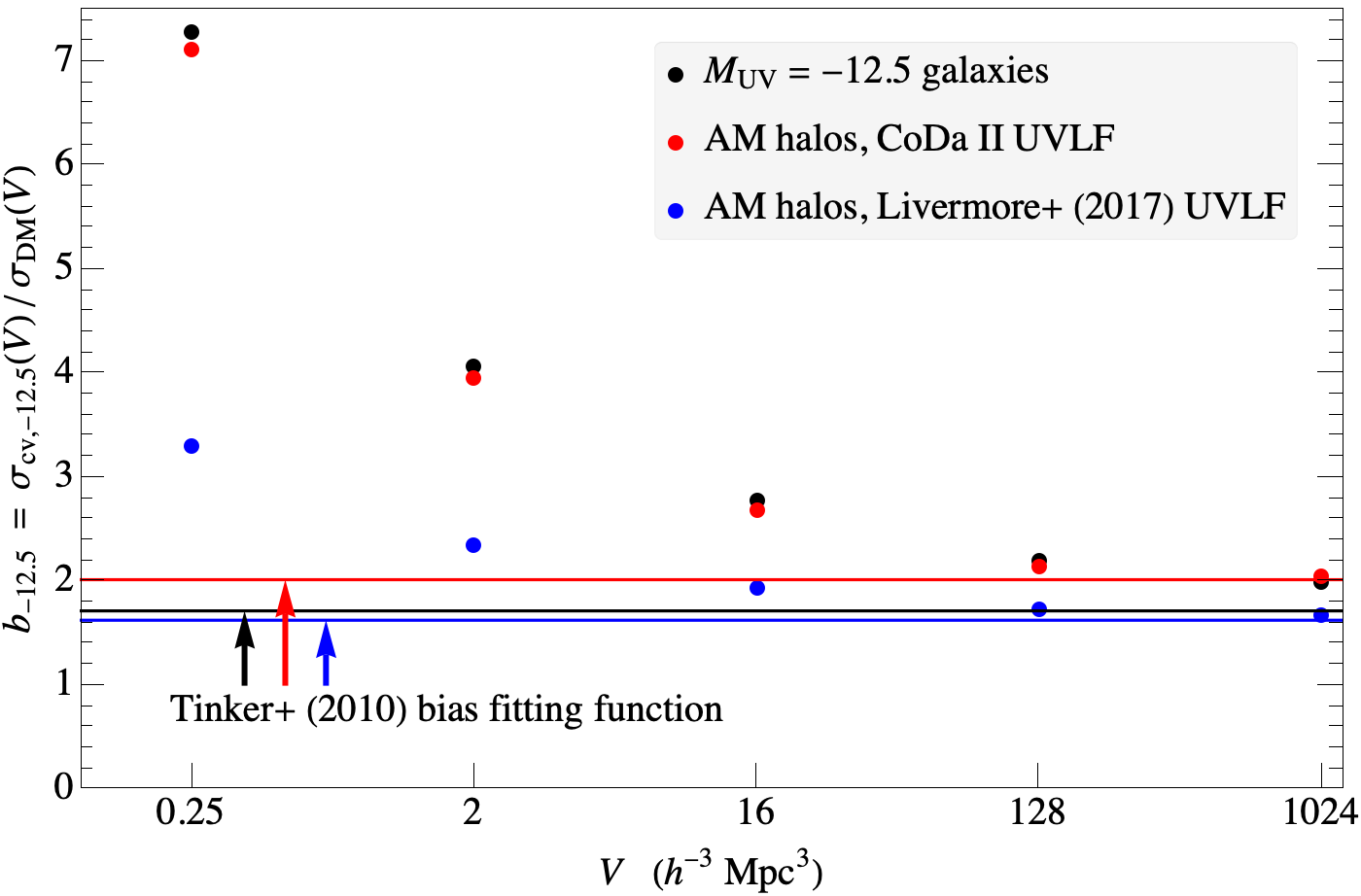}
    \caption{The bias of $M_\text{UV} = -12.5$ galaxies in CoDa~II at $z=6$, computed by counting galaxies/halos in volumes of size $V$. 
    The smallest volume shown, $V = 0.25~h^{-3}$~cMpc$^3 = 0.8$~cMpc$^3$, is similar to the effective volume searched by \citet{Livermore17} at this magnitude, while the largest volume shown, $V = 1024~h^{-3}$~cMpc$^3 = 3300$~cMpc$^3$, corresponds to our subvolumes.
    The black points are obtained by counting galaxies in the magnitude bin $M_\text{UV} = -12.5 \pm 0.25$ directly, whereas the red and blue points are obtained by counting halos with masses obtained by abundance matching (AM) to the magnitude bin, using the CoDa II HMF and the UVLFs from CoDa II and \citet{Livermore17}, respectively.
    For comparison, we show the bias as computed using the fitting function from \citet{Tinker10} as horizontal lines. The black line is the bias obtained by abundance matching using the HMF from \citet{Tinker08} and the UVLF from \citet{Livermore17}. The red line is obtained by abundance matching using the CoDa~II HMF and UVLF, which is analogous to the red points. The blue line is obtained by abundance matching using the CoDa~II HMF and \citet{Livermore17} UVLF, which is analogous to the blue points. Our results diverge from the \citet{Tinker10} fitting function estimates at small volumes.}
    \label{fig:bias}
\end{figure}

\begin{figure}
    \centering
    \includegraphics[width=\columnwidth]{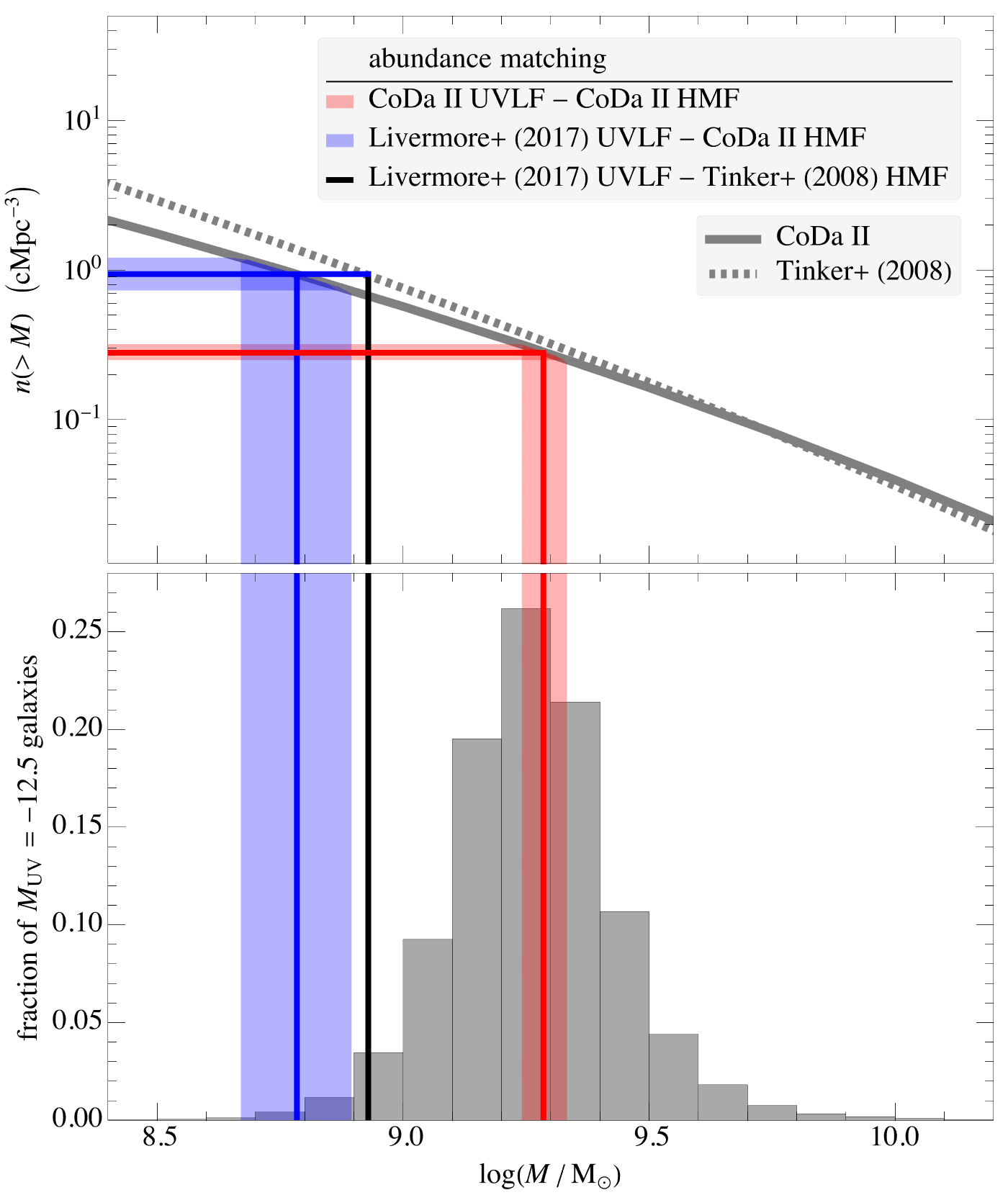}
    \caption{Top: Cumulative number densities of halos with mass $> M$ for CoDa~II (solid gray) compared to that obtained from the HMF of \citet{Tinker08} (dashed gray), both of which are used for abundance matching. 
    The horizontal red and blue bands represent the cumulative number densities of galaxies with magnitudes brighter than the bin $-12.5 \pm 0.25$, according to the UVLFs from CoDa~II and \citet{Livermore17}, respectively.
    The vertical red and blue bands represent the mass ranges obtained by matching the aforementioned galaxy number densities with the CoDa~II cumulative halo number density curve.
    These mass ranges are used to compute the red and blue points in Fig.~\ref{fig:bias}.
    The red and blue lines represent abundance matching for the center of the magnitude bin ($M_\text{UV}<-12.5$), and the resulting masses are used to obtain the red and blue horizontal lines in Fig.~\ref{fig:bias}.
    The black vertical line is obtained by instead matching the \citet{Livermore17} abundance of $M_\text{UV}<-12.5$ galaxies to the \citet{Tinker08} cumulative halo number density curve, and the resulting mass is used to obtain the black horizontal line in Fig.~\ref{fig:bias}.
    Bottom: The distribution of halo masses for $M_\text{UV}=-12.5 \pm 0.25$ galaxies in CoDa~II, which is encoded in the black points in Fig.~\ref{fig:bias}, compared to the abundance-matched masses discussed above.}
    \label{fig:abundance-matching}
\end{figure}

To illustrate this, we further sub-divide the CoDa~II simulation at $z=6$ into sets of `sub-subvolumes' of different scales, all the way down to $V = 0.8$~cMpc$^3$, which is close\footnote{We obtained the sub-subvolumes by evenly sub-dividing the 3300 cMpc$^3$ subvolumes into a sequence of smaller volumes with integer numbers of grid cells per dimension, and arrived at $3300/4096 = 0.8$ cMpc$^3$ as the closest sub-division to $V_\text{eff} = 0.73$ cMpc$^3$.} to the effective volume searched by \citet{Livermore17} at $M_\text{UV} = -12.5$.
We then computed the bias of galaxies in the magnitude bin $M_\text{UV} = -12.5\pm0.25$ in each of these sets of sub-subvolumes by computing the variance in their number densities to obtain $\sigma_\textsc{cv,l}$, and show the results as a function of volume in Fig.~\ref{fig:bias} (black points).
For comparison, we show the bias as estimated from the (large-scale) analysis of \citet{Tinker10} for galaxies at this magnitude as a horizontal black line.
We obtained this estimate by applying the \citet{Tinker10} bias fitting function to the halo mass obtained by abundance matching the \citet{Tinker08} HMF with the \citet{Livermore17} UVLF at $M_\text{UV} = -12.5$.
As expected, our bias estimates from CoDa~II approach the \citet{Tinker10} estimate on large scales, but deviate substantially from the latter on smaller scales.
In particular, near the effective volume searched by \citet{Livermore17} at $M_\text{UV} = -12.5$, our bias ($b_{-12.5}$) is larger than the (large-scale) \citet{Tinker10} estimate by a factor of { 4.2}.
Correspondingly, our estimate of the contribution to the uncertainty due to cosmic variance from an observation at this volume and magnitude ($\sigma_\textsc{cv,-12.5}$) is a factor of { 4.2} larger than that which \cite{Robertson14} (and, following the former, \citealp{Livermore17}) would have found.

{ This result compares the variance of galaxies in a given \textit{luminosity} range in the CoDa~II simulation to that of halos of a given \textit{mass} according to the scale-independent bias estimate from \citet{Tinker10}.}
In order to perform a more `apples-to-apples' comparison (i.e. `mass-to-mass', rather than `luminosity-to-mass'), we also computed the bias in two different ways, for both our simulation results and the \citet{Tinker10} estimate.
For our simulation results, rather than calculating the variance in the number density of $M_\text{UV} = -12.5\pm0.25$ galaxies directly, we instead calculate the variance in halos with masses in a range obtained by abundance matching to this magnitude bin, i.e.
\begin{equation}
    \sigma_\textsc{cv,m}(V) = \frac{ \sqrt{ \langle (n_\textsc{m}(\bm{x},V)-\Bar{n}_\textsc{m})^2\rangle_{\bm{x}} } }{\Bar{n}_\textsc{m}} = b_\textsc{m}(V) \, \sigma_\textsc{dm}(V)
\end{equation}
where the subscript $M$ now refers to the same quantities as before -- uncertainty due to cosmic variance, local and global number densities, and bias -- but for halos with mass $M$, rather than galaxies with a given luminosity.
We perform the abundance matching using our numerical CoDa~II HMF and either the CoDa~II UVLF (red points) or the \citet{Livermore17} UVLF (blue points).
Then, we apply these abundance-matched halo masses to the \citet{Tinker10} fitting function to obtain new estimates, which are shown in Fig.~\ref{fig:bias} as red and blue horizontal lines, to be compared with the red and blue points, respectively.
We illustrate the relationship between these different abundance matching methods in Fig.~\ref{fig:abundance-matching}.

As before, the points converge to the \citet{Tinker10} estimate on large scales, but diverge on small scales.
{  Since the red points show the cosmic variance in halos that are self-consistently abundance-matched to the galaxies represented by the black points, they naturally exhibit roughly the same behavior as the black points, and are a factor of 3.5 times higher than the corresponding \citet{Tinker10} estimate at $V =0.25 \,h^{-3}\, {\rm cMpc}^3$.}
Although the blue points (halos chosen by abundance matching the CoDa~II HMF to the \citealp{Livermore17} UVLF, rather than the CoDa~II UVLF) exhibit less of a discrepancy, our CoDa~II result is still a factor of 2 larger than the comparable fitting function estimate on this small scale.
Thus, we find that the procedure adopted by \citet{Livermore17} to model their uncertainty due to cosmic variance
underestimates the contribution from their observation of a $M_\text{UV} = -12.5$ galaxy in an effective volume of 0.73 cMpc$^3$ by at least a factor of 2.

For future small-effective-volume searches that require cosmic variance estimates, we suggest computing the variance directly from simulations, on the scale of the effective search volume, as we have done here, rather than using fitting functions that are only applicable in the large-scale limit.
Even using dark-matter-only simulations for this purpose will provide a much more accurate estimate of the uncertainty due to cosmic variance than the latter approach.
As we discussed in the introduction and \S\ref{sec:patchy}, cosmic variance in the UVLF is a result of both variance in the HMF and variance in the SFRs of halos of a given mass.
While dark-matter-only simulations cannot account for the latter, the proximity of the red (variance in halos of a given mass) and black (variance in galaxies of a given luminosity) points in Fig.~\ref{fig:bias} indicates that the variance in SFRs is a sub-dominant contributor to the deviation of our result from the large-scale fitting function, since this variance is accounted for in the black points but not the red ones.
Thus, most of the deviation is captured just by accounting for variance in dark matter halo abundances on small scales, which can be approximated from dark-matter-only simulations.

\subsection{Fitting Functions for the CoDa~II UVLF}
\label{sec:fit}

\begin{figure}
    \centering
    \includegraphics[width=\columnwidth]{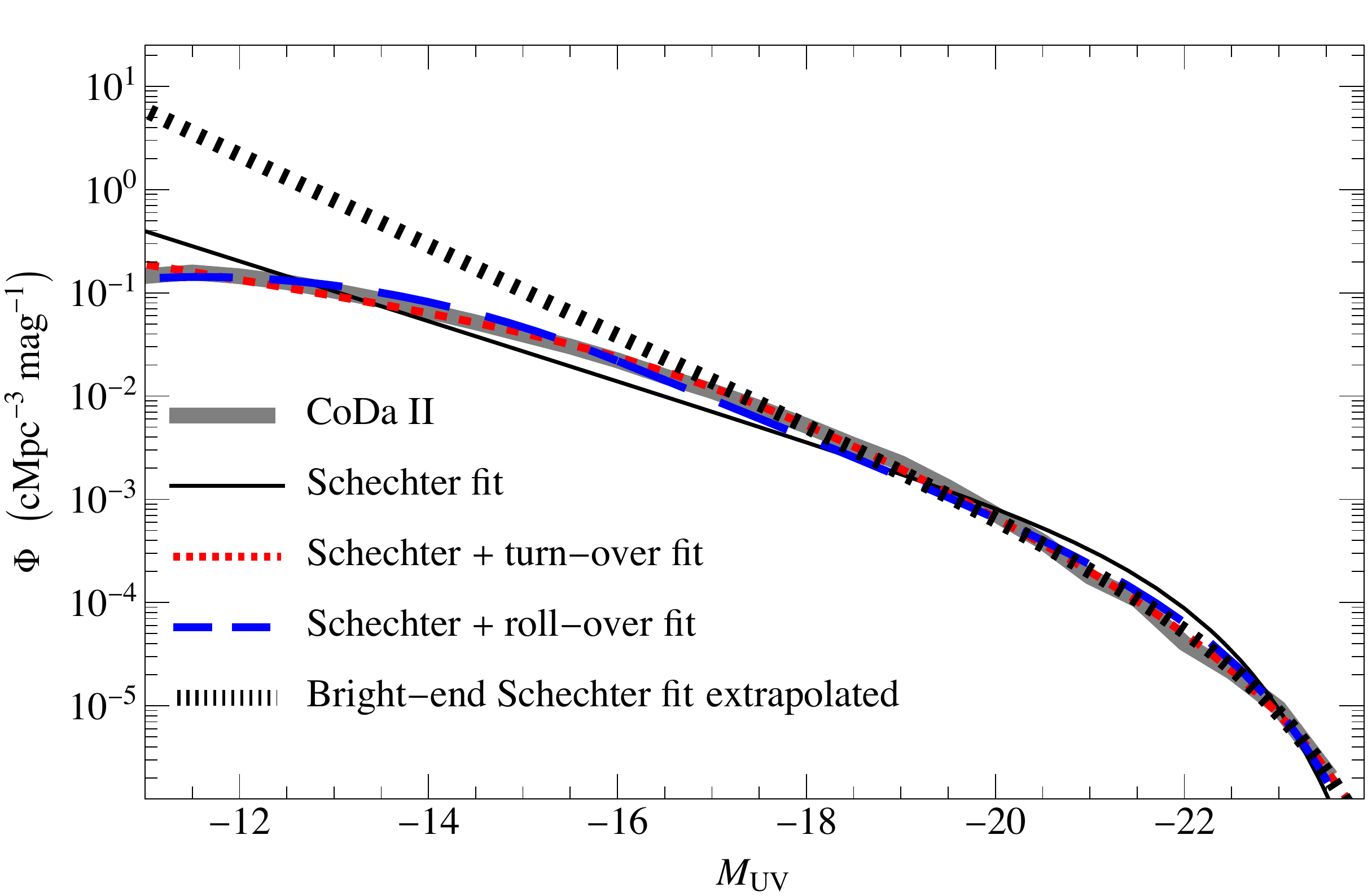}
    \includegraphics[width=\columnwidth]{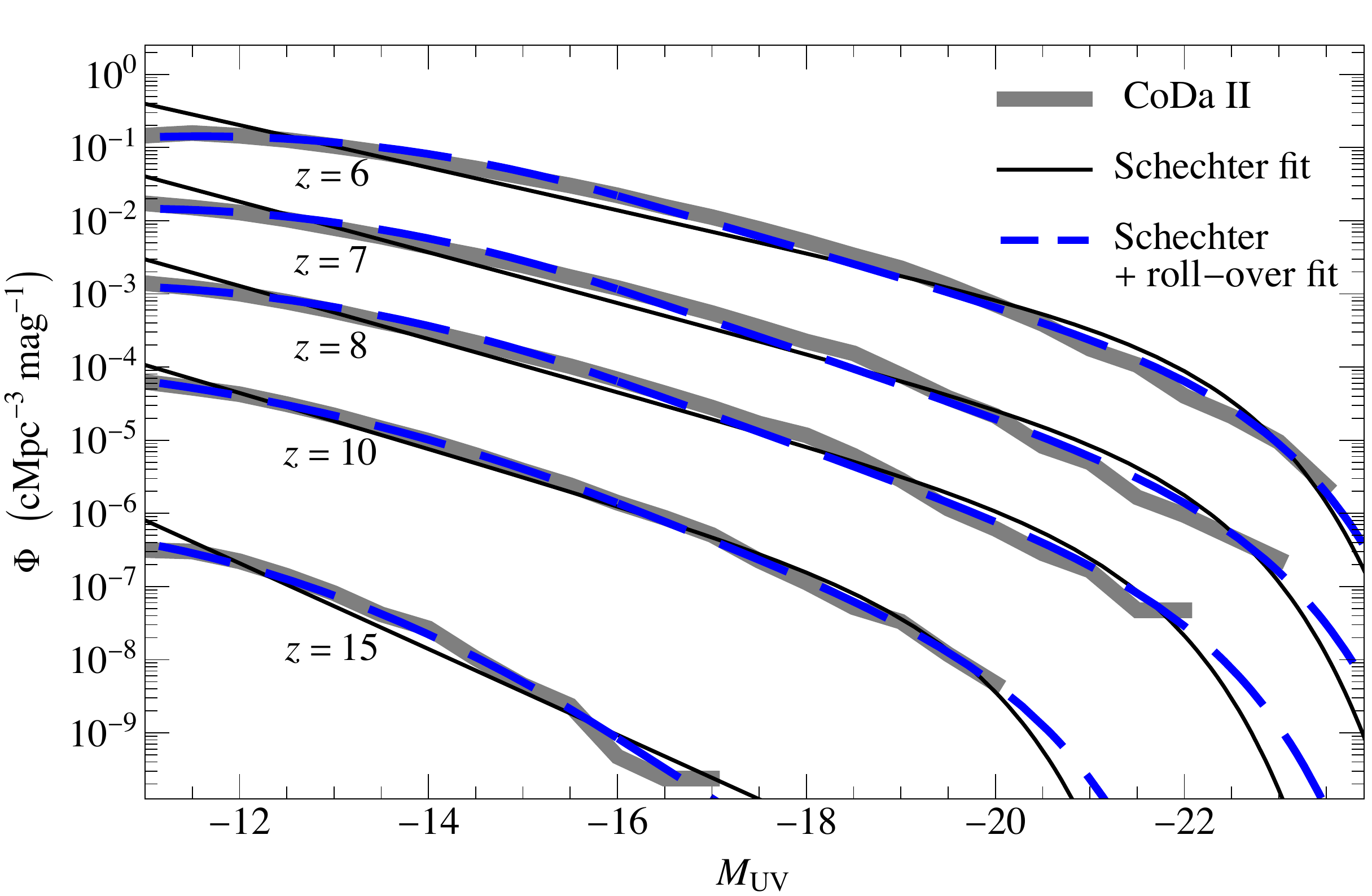}
    \caption{Left: A comparison of various fitting functions to the CoDa~II full volume UVLF, as labeled. The black hashed line is a Schechter function fit only to magnitudes bright-ward of $-17$, but extrapolated faint-ward from there, to illustrate the degree to which the CoDa~II UVLF deviates from the faint-end power-law behavior of the Schechter function.
    Right: Pure Schechter and Schechter+roll-over fits at different redshifts. Curves for $z=7,8,10,15$ are shifted downward by 1, 2, 3, and 4 dex, respectively.}
    \label{fig:fit}
\end{figure}

To parameterize the shape of our CoDa~II UVLFs, it is useful to fit them to Schechter functions, or modifications thereof. 
The Schechter function is defined as
\begin{equation}
    \Phi_\textsc{l} = \left(\frac{\Phi_*}{L_*}\right)\left(\frac{L}{L_*}\right)^\alpha \exp\left(-\frac{L}{L_*}\right)
\end{equation}
where $\Phi_*$, $L_*$, and $\alpha$ are free parameters, the last of which is the logarithmic slope of the faint-end.
By convention, the Schechter function is usually reparameterized in terms of magnitude as
\begin{equation}
    \Phi_\textsc{m} = 0.4 \ln(10) \Phi_* 10^{0.4(M_* - M_\text{UV})\!(\alpha+1)} \!\exp\!\left[ -10^{0.4(M_* - M_\text{UV})} \!\right]
\end{equation}
We fit this function to our $z=6$ CoDa~II UVLF in the top panel of Fig.~\ref{fig:fit} (thin black line). 
As can be seen there, the shape of the Schechter function fails to capture the curvature of the CoDa~II UVLF, due to the fact that the Schechter function maintains a constant power law slope at the faint end, while the CoDa~II UVLF flattens out gradually towards fainter magnitudes, due to the suppression of star formation in low-mass halos caused by reionization feedback.
We can construct a better fit by allowing the power law slope to change above a certain magnitude, using what we will refer to as the Schechter + turn-over function from \citet{Jaacks13}:
\begin{equation}
    \Phi_\textsc{t} = \Phi_\textsc{m} \left( 1+10^{0.4(M_T - M_\text{UV})\beta} \right)^{-1}
\end{equation}
where $M_T$ is the so-called turn-over magnitude, above which the power law slope changes (in luminosity space) from $\alpha$ to $\alpha-\beta$.
The fit of this function to the CoDa~II UVLF is shown as the red dashed line in Fig.~\ref{fig:fit}.
The fitted model exhibits a turn-over at
\begin{equation}
    M_T = -16.9 \pm 0.4
\end{equation}
from a power law slope of
\begin{equation}
    \alpha = -2.23 \pm 0.05
\end{equation}
to
\begin{equation}
    \alpha - \beta = -1.36 \pm 0.06
\end{equation}
Given this turn-over magnitude, we can see more clearly the deviation of our CoDa~II UVLF from the Schechter function, by fitting only the bright-end of the UVLF ($M_\text{UV} < -17$) to a Schechter function, and extrapolating that bright-end fit faint-ward.
This is shown as the black hashed line in the figure.
We can see that the bright-end fit clearly starts to deviate from the faint-end of the UVLF for $M_\text{UV} \gtrsim -16$.

\begin{figure*}
    \centering
    \includegraphics[width=\textwidth]{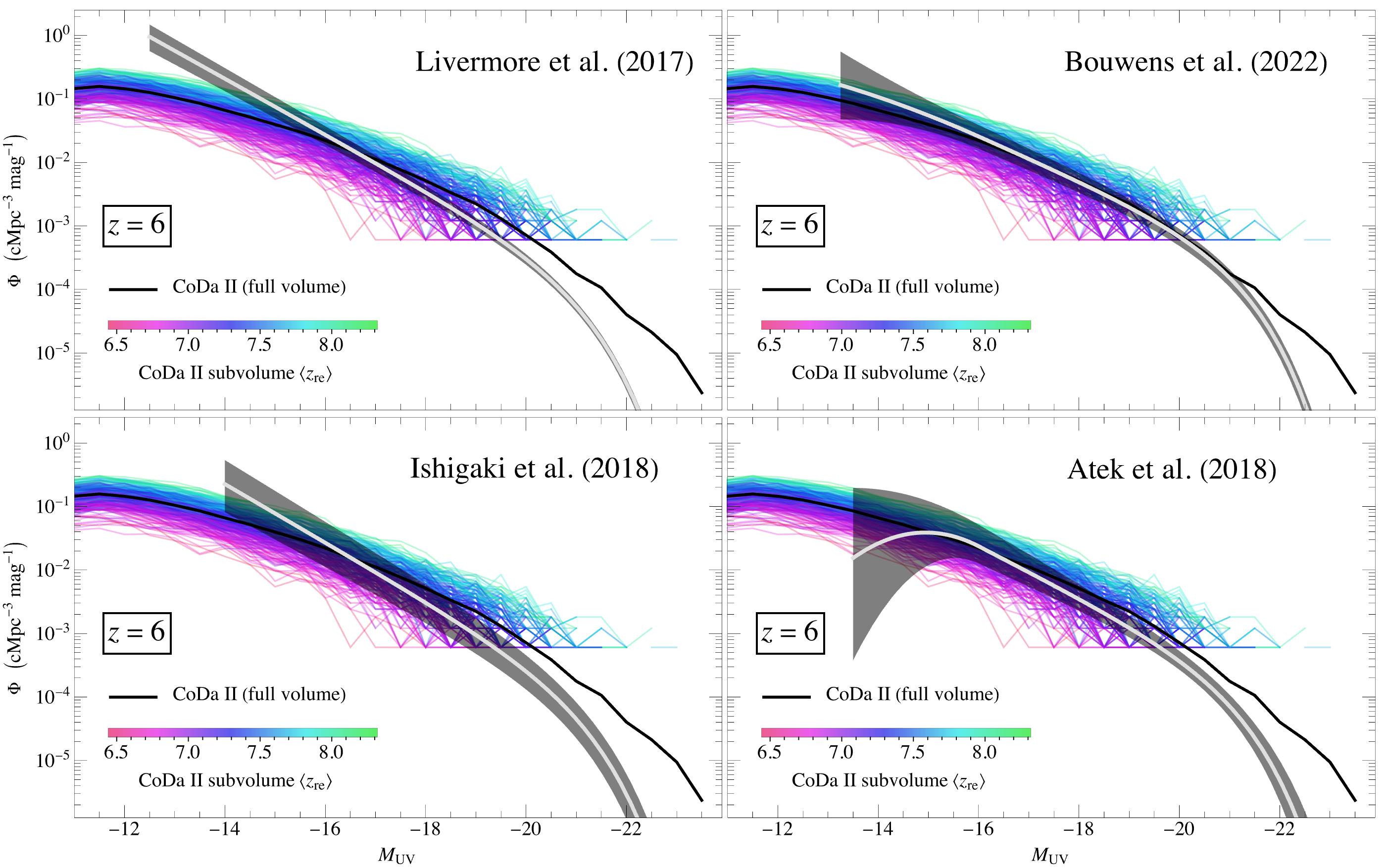}
    \caption{Comparison of our $z=6$ CoDa~II UVLF with the fitting functions (light gray lines with errors shown in dark shading) from \citet{Livermore17,Bouwens22,Ishigaki18,Atek18}. \citet{Livermore17,Ishigaki18} prefer pure Schechter function fits, while \citet{Bouwens22,Atek18} prefer Schechter functions modified to roll-over at the faint end. Our CoDa~II results favor the latter.}
    \label{fig:obs}
\end{figure*}

Another modification of the Schechter function was proposed by \citet{Bouwens17}, wherein the faint end gradually `rolls' over at magnitudes fainter than $-16$, taking on a smoothly-varying parabolic shape, rather than another power law. 
We will refer to this as the Schechter + roll-over function, which is defined as
\begin{equation}
    \Phi_\textsc{r} =
    \begin{cases}
        \Phi_\textsc{m}, & M_\text{UV} < -16 \\
        \Phi_\textsc{m} 10^{-0.4(-16-M_\text{UV})^2 \delta} , & M_\text{UV} > -16
    \end{cases}
\end{equation} 
The advantage of this fitting function is that it captures the continued change in slope that occurs in the UVLF towards fainter magnitudes.
We show this fit as the blue dashed line in Fig.~\ref{fig:fit}.
In addition to $z=6$, we also show the pure Schechter and Schechter + roll-over fits to the CoDa~II UVLF at $z = 7,8,10,15$ in the bottom panel of Fig.~\ref{fig:fit}.
With the exception of $z=15$ (where our data is most limited), the pure Schechter fit gets worse with decreasing $z$, due to the increasing suppression of low-mass halos as reionization progresses.

Amongst the previously discussed HFF analyses, \citet{Livermore17} and \citet{Ishigaki18} preferred pure Schechter function fits, as their data showed no sign of a turn-over at the faint-end, while \citet{Bouwens22} and \citet{Atek18} preferred Schechter + roll-over fits.
We show each of their best-fit functions, with corresponding error, compared to our CoDa~II UVLF in Fig.~\ref{fig:obs}, and provide the best-fit parameter estimates in Table~\ref{tab:comp}.
As can be seen, CoDa~II has a clear preference for the roll-over models, with parameters that are most similar to those of \citet{Bouwens22}.

\renewcommand{\arraystretch}{1.25}
\begin{table*}
    \centering
    \caption{Best-fit (modified) Schechter function parameters for CoDa~II's globally-averaged UVLFs vs. those inferred from HFF data by 4 studies. The HFF results are for $z=6$.}
    \begin{tabular}{l|c c c c}
        \hline
         & $\log \Phi_{*}$ & $M_{*}$ & $\alpha$ & $\delta$ \\
        \hline
        \cite{Livermore17} & $-3.647 \substack{+0.039 \\ -0.037}$ & $-20.825 \substack{+0.055 \\ -0.043}$ & $-2.10 \substack{+0.05 \\ -0.04}$ & -- \\
        \cite{Bouwens22} & $-3.24 \pm 0.08$ & $-20.87 \pm 0.07$ & $-1.87 \pm 0.04$ & $0.05 \pm 0.10$ \\
        \cite{Ishigaki18} & $-3.78 \substack{+0.15 \\ -0.15}$ & $-20.89 \substack{+0.17 \\ -0.13}$ & $-2.15 \substack{+0.08 \\ -0.06}$ & -- \\
        \cite{Atek18} & $-3.54 \substack{+0.06 \\ -0.07}$ & $-20.84 \substack{+0.27 \\ -0.30}$ & $-2.01 \substack{+0.12 \\ -0.14}$ & $0.48 \substack{+0.49 \\ -0.25}$ \\
        CoDa~II ($z=6$) & $-3.95 \pm 0.086$ & $-22.3 \pm 0.099$ & $-1.92 \pm 0.026$ & $0.105 \pm 0.011$\\
        \hline
        CoDa~II ($z=7$) & $-4.60 \pm 0.16$ & $-22.3 \pm 0.21$ & $-2.08 \pm 0.039$ & $0.106 \pm 0.016$\\
        CoDa~II ($z=8$) & $-4.74 \pm 0.22$ & $-21.6 \pm 0.30$ & $-2.14 \pm 0.046$ & $0.100 \pm 0.016$\\
        CoDa~II ($z=10$) & $-4.67 \pm 0.11$ & $-19.8 \pm 0.13$ & $-2.23 \pm 0.026$ & $0.080 \pm 0.007$\\
        CoDa~II ($z=15$) & $-6.50 \pm 0.08$ & $-17.8 \pm 0.07$ & $-3.06 \pm 0.147$ & $0.146 \pm 0.034$\\
        \hline
    \end{tabular}
    \label{tab:comp}
\end{table*}
\renewcommand{\arraystretch}{0.8}

\section{Conclusions}
\label{sec:conclusion}

In this work, we analyzed the CoDa II simulation to study the spatial and temporal variations in the high-redshift UVLF during the EOR.
We find that the UVLF is strongly correlated with the local reionization history of the region in which it is measured.
Earlier-reionizing regions, which have higher overdensities and HMFs, have correspondingly higher UVLFs with brighter exponential cut-off magnitudes ($M_*$).
Therefore, the results of small-volume observations of the high-$z$ UVLF, e.g. those made through high-magnification gravitational lenses, will depend on where one looks (e.g. at an early-, intermediate-, or late-reionizing patch of the universe).
In addition, such observations will also depend on \textit{when} one looks -- i.e. before or after the observed region has been reionized.
Due to the fact that the photoheating of gas in the IGM during reionization suppresses the formation of stars in low-mass halos, the faint-end of the UVLF in a given region evolves over the course of its local reionization, becoming increasingly flattened over time.
The UVLF of a region observed after its local reionization will exhibit a turn-over or roll-over at the faint-end, at magnitudes $\gtrsim -17$.
By $z=6$, when most of the universe is reionized, the global UVLF exhibits this faint-end turn-over, as well, and the gradual flattening of the global faint-end slope over time can be seen starting from $z=10$.

We find that our CoDa II UVLFs are in good agreement with data from HFF lensing surveys, with the exception of a single data point from \citet{Livermore17} of a $M_\text{UV}=-12.5$ galaxy observed at $z=6$ in an effective search volume of $V_\text{eff} = 0.73$~cMpc$^3$. 
This observation implies a UVLF that is $\sim10$ times higher at this magnitude than what we predict from CoDa~II, as well as a faint-end slope that remains steep down to $z=6$.
This motivated us to ask about the variation of the UVLF with position in CoDa~II, to determine the likelihood of the faint-end detection reported by \citet{Livermore17} given the limitations of the search technique involving gravitational lensing amplification by a foreground cluster.
Given the abundance of $M_\text{UV}=-12.5$ galaxies in our CoDa~II simulation, the probability of encountering one in a 
randomly-placed 0.73 cMpc$^3$ search volume at that redshift is found to be relatively small, at $\sim 4.5\%$.  As such, our results are more consistent with the analysis of \citet{Bouwens17}, wherein this observation is re-interpreted as a brighter galaxy in a larger effective search volume, with a greater uncertainty attributed to the lensing models.
Indeed, of the four observational papers whose UVLFs we compare to our simulated one, our results are most similar to those of \citet{Bouwens22}.

Furthermore, given the stark differences in galaxy abundances on small scales resulting from spatial and temporal variations in density and reionization history, we believe that the uncertainty due to cosmic variance attributed to these observations -- when using them to infer the global UVLF -- has been underestimated.
{  The uncertainty due to cosmic variance for lensing surveys like the HFFs is typically estimated according to \citet{Robertson14}, in which it is  proportional to the galaxy clustering bias derived from the large-scale $N$-body simulations and halo bias analysis of \citet{Tinker10}.}
However, for these small-scale lensing observations, we find that the bias of $M_\text{UV}=-12.5$ galaxies in 0.8 cMpc$^3$ volumes is a factor of 2--{ 4} higher than this large-scale estimate, primarily due to the increased variance in dark matter halo abundances on small scales vs. large scales.
As next-generation space- and ground-based telescopes start to probe the high-$z$ Universe down to fainter magnitudes than ever before, we expect a more thorough accounting of cosmic variance -- one that accommodates the full scope of spatial and temporal inhomogeneity during the EOR -- to be essential for reconciling competing observational inferences and theoretical models.

\section*{Acknowledgements}

This material is based upon work supported by the National Science Foundation Graduate Research Fellowship Program under Grant No. DGE-1610403. Any opinions, findings, and conclusions or recommendations expressed in this material are those of the authors and do not necessarily reflect the views of the NSF. 
This research used resources of the Oak Ridge Leadership Computing Facility at the Oak Ridge National Lab, which is supported by the Office of Science of the US Dept.~of Energy under Contract No.~DE-AC05-00OR22725. CoDa~II was performed on Titan at OLCF under DOE INCITE 2016 award to Project AST031.
Data analysis was performed using the computing resources from NSF XSEDE grant TG-AST090005 and the Texas Advanced Computing Center at the University of Texas at Austin.
PRS acknowledges support from NASA under Grant No. 80NSSC22K175.
JL acknowledges support from the DFG via the Heidelberg Cluster of Excellence STRUCTURES in the framework of Germany’s Excellence Strategy (grant EXC-2181/1-390900948).
JS acknowledges support from the ANR LOCALIZATION project, grant ANR-21-CE31-0019 of the French Agence Nationale de la Recherche.
KA is supported by Korea NRF-2021R1A2C1095136, NRF-2016R1A5A1013277 and RS-2022-00197685. KA also appreciates the APCTP for its hospitality during completion of this work.
ITI was supported by the Science and Technology Facilities Council [grant numbers ST/I000976/1 and ST/T000473/1] and the Southeast Physics Network.
HP was supported by the World Premier International Research Center Initiative, MEXT, Japan and JSPS KAKENHI grant No. 19K23455.
GY acknowledges financial support from MICIU/FEDER under project grant PGC2018-094975-C21.

%\vspace{-2em}

% %%%%%%%%%%%%%%%%%%%%%%%%%%%%%%%%%%%%%%%%%%%%%%%%%%
\section*{Data Availability}

The data underlying this article are available in the article and in its online supplementary material.

%\vspace{-2em}

%%%%%%%%%%%%%%%%%%%% REFERENCES %%%%%%%%%%%%%%%%%%

% The best way to enter references is to use BibTeX:

\bibliographystyle{mnras}
\bibliography{ref}

% Alternatively you could enter them by hand, like this:
% This method is tedious and prone to error if you have lots of references
%\begin{thebibliography}{99}
%\bibitem[\protect\citeauthoryear{Author}{2012}]{Author2012}
%Author A.~N., 2013, Journal of Improbable Astronomy, 1, 1
%\bibitem[\protect\citeauthoryear{Others}{2013}]{Others2013}
%Others S., 2012, Journal of Interesting Stuff, 17, 198
%\end{thebibliography}

%%%%%%%%%%%%%%%%%%%%%%%%%%%%%%%%%%%%%%%%%%%%%%%%%%

%%%%%%%%%%%%%%%%% APPENDICES %%%%%%%%%%%%%%%%%%%%%

\appendix
{ 
\section{Comparison of the CoDa II Global Halo Mass Function to Standard Fits From Dark-Matter-Only N-body Results}
A comparison of the global CoDa II HMF at $z=6$ to the fitting functions of \citet{SMT01,Tinker08,Watson13}, derived from dark-matter-only (``DMO'') $N$-body simulations, is shown in Fig.~\ref{fig:HMF_comp}.
    The top panel shows the HMFs themselves, while the bottom panel shows the log of the ratios of the CoDa II HMF to each of these fitting functions, as labeled. 
    The shaded region in the bottom panel roughly highlights the trend in the difference between CoDa II and the DMO simulation fits.  There is a tendency for the CoDa II HMF to be a bit lower than the DMO fits over a broad range of masses, by an amount that is comparable to the spread in amplitudes of these different fits. This trend steepens, however, at the low-mass end, below $M \lesssim 10^{8.5}~\text{M}_\odot$. There, we approach the resolution limit of our simulation, and the trend for the ratios is that of a power-law with a relatively steep index of $\sim 0.35$.
    Above this mass, however, where CoDa II halos contain more than $\sim 1000$ DM particles, the trend flattens to an index of $\sim$ 0.1--0.15.
    This latter power-law continues over orders of  magnitude in halo mass, all well above our resolution limit, until around $M\sim 10^{11}~\text{M}_\odot$.
    Therefore, we believe this lowering of the HMF in CoDa II relative to the DMO fits is a physical consequence of including hydrodynamics in the simulation, rather than a numerical limitation.
In short, since our results here for the inhomogeneity of the UVLF, including our use of the CoDa~II HMF for abundance-matching down through the faint end of the LF, only depend on the HMF above $M \gtrsim 10^{8.5}~\text{M}_\odot$, they should be robust with respect to our numerical resolution of the HMF.   

As shown here by analyzing the CoDa II simulation, the self-consistent treatment of halo formation including baryonic feedback effects from star formation, supernovae, and reionization tends to reduce the HMF relative to that predicted from DMO simulations by a modest amount. 
This trend was demonstrated previously by \citet{Sawala13}, by directly comparing their GIMIC hydrodynamical simulations with DMO simulations from the same initial conditions, for halos all the way up to $M \lesssim 10^{12}~\text{M}_\odot$. 
As these authors reported, while the two types of simulation agreed well on large scales, objects below this mass scale had systematically lower masses in the GIMIC simulation (i.e. with baryonic hydrodynamics) than in the DMO simulation, resulting in a corresponding shift downward in the HMF, by a larger amount for smaller mass halos. This is consistent with the results found here for CoDa II.  In fact, the CoDa II simulation strengthens the case for this trend, since it is based on fully-coupled radiation-hydrodynamics (i.e. with radiative transfer), while the GIMIC simulation (with coarser particle-mass resolution than ours but comparable length resolution)  adopted a uniform, optically-thin photoionizing background, instead.

\begin{figure}
    \centering
    \includegraphics[width=\columnwidth]{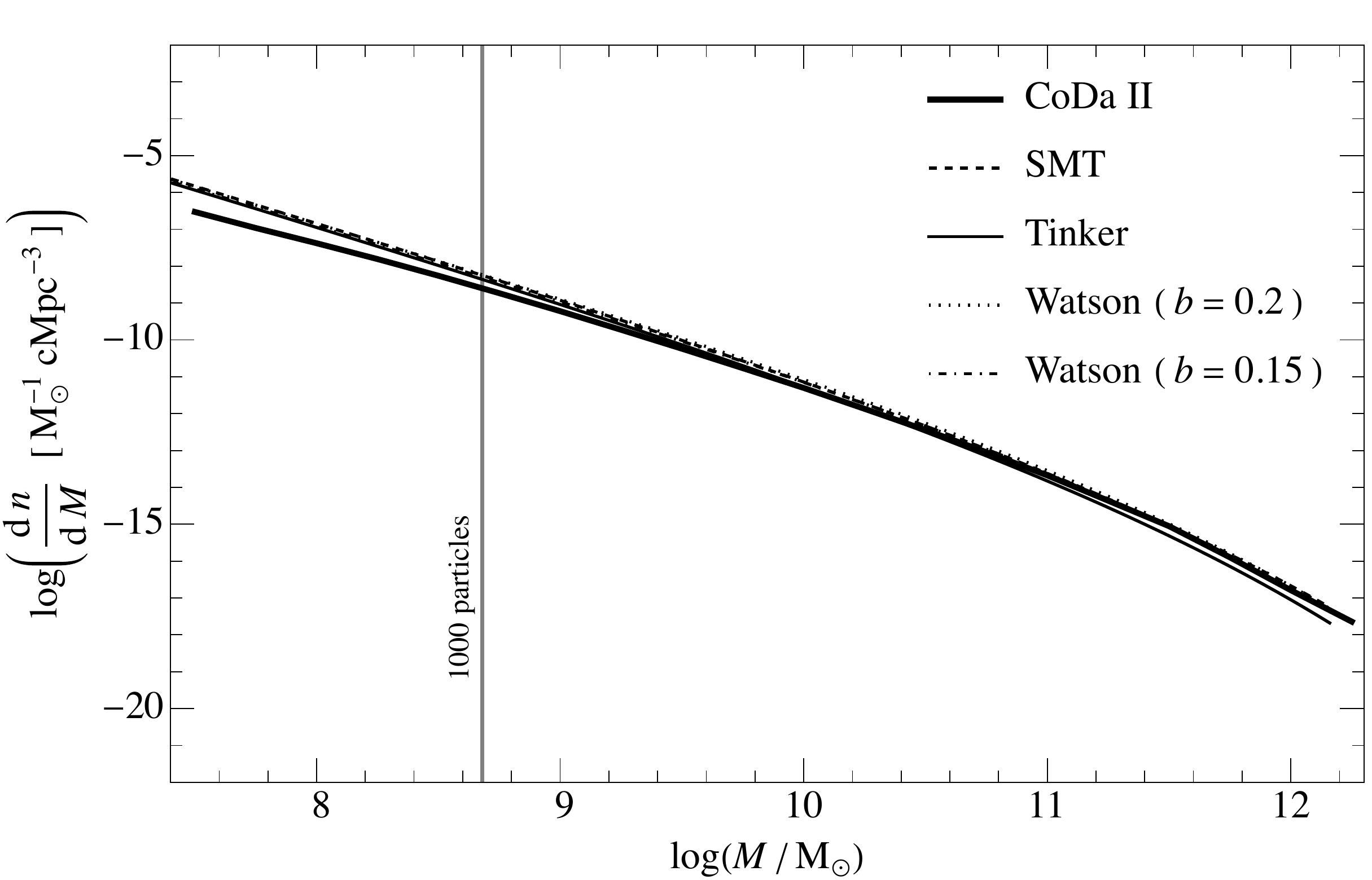}\\
    \quad\includegraphics[width=\columnwidth]{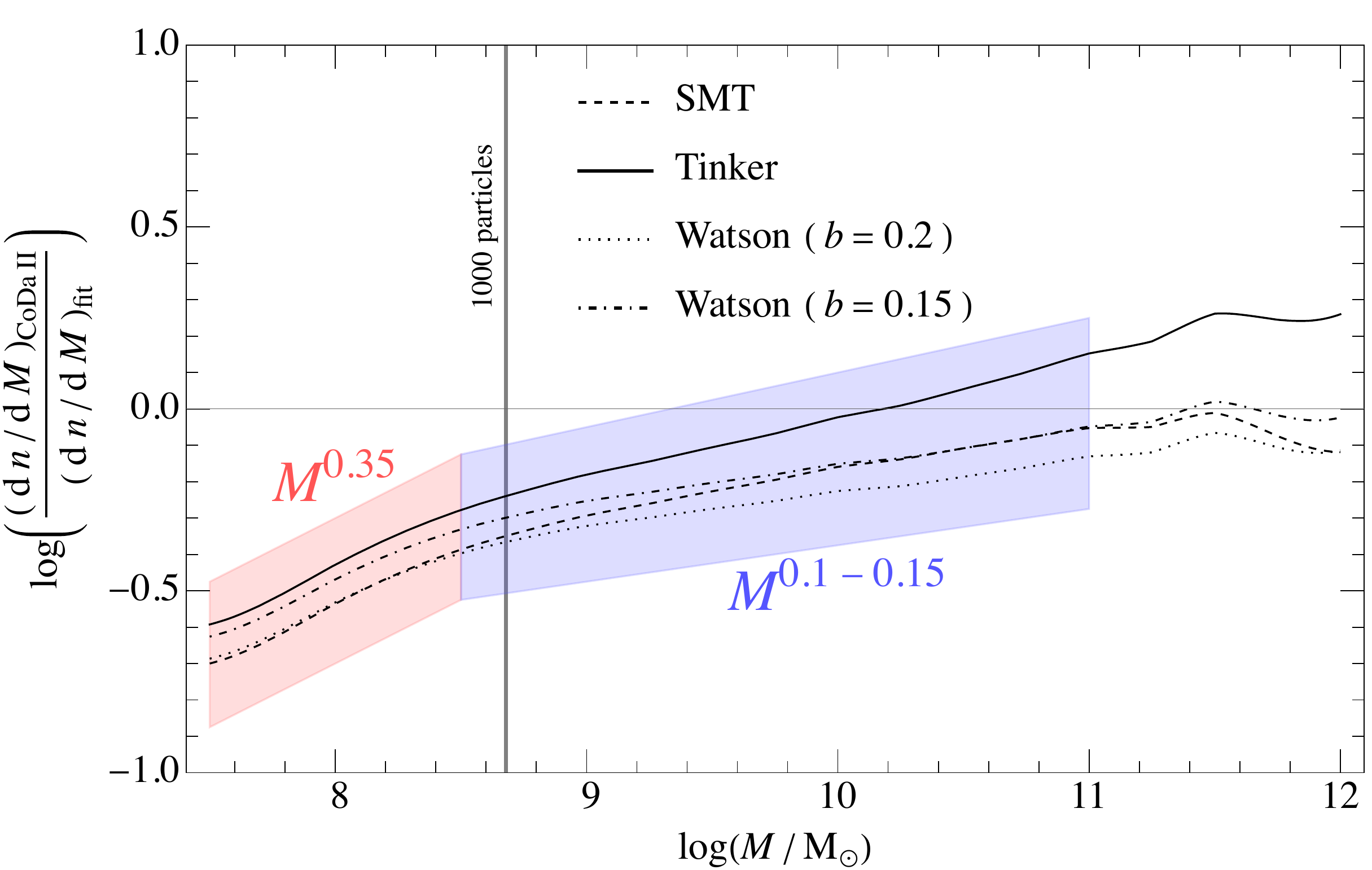}
    \caption{  A comparison of the CoDa II HMF at $z=6$ to the fitting functions from \citet{SMT01,Tinker08,Watson13}.
    We show two versions of the \citet{Watson13} fit, adopting Friends-of-Friends linking parameters of $b=0.2$ and $b=0.15$.
    The top panel shows the HMFs themselves, while the bottom panel shows the log ratio of the CoDa II HMF to the various fitting functions, as labeled.
    The shaded region in the bottom panel roughly highlights the trend in the difference between CoDa II and the dark-matter-only simulation fits.
    At the low-mass end ($M \lesssim 10^{8.5}~\text{M}_\odot$), we approach the resolution limit of our simulation, and the trend is that the ratios follow a power-law with an index of $\sim 0.35$.
    Above this mass, however, the trend flattens to an index of $\sim 0.1-0.15$.
    This latter power-law continues well above our resolution limit, until around $M\sim 10^{11}~\text{M}_\odot$.
    Therefore, we believe the latter difference is a physical consequence of including hydrodynamics in the simulation, rather than a numerical limitation.}
    \label{fig:HMF_comp}
\end{figure}

\section{Related Measures of the UVLF}

We present here some additional quantities related to the UVLF, which may be useful for further theoretical or observational comparisons. 
In Fig.~\ref{fig:UVLFnum}, we plot the number of galaxies in CoDa II that fall within each magnitude and $z_\text{re}$ bin used throughout this paper.
This figure amounts to a renormalization of the various curves shown in Fig.~\ref{fig:UVLF}, to show the relative contributions of early-, intermediate-, and late-reionizing regions to the global UVLF.
For instance, one feature that can be gleaned easily from this figure is that the magnitude at which our early-reionization bin transitions from the dominant contributor to a sub-dominant contributor becomes brighter over time.
This is due to the fact that while the early-reionizing regions are relatively overdense, they also occupy a relatively small volume.
Therefore, these regions form the dominant share of bright galaxies at early times, but are eventually over-taken as the less-overdense-but-larger-volume regions become increasingly non-linear at later times.  

This plot of the actual numbers of galaxies in the CoDa II simulation volume at each redshift also illustrates why a simulation with as large a volume as CoDa II is required in order to make a statistically-meaningful analysis of the UVLF possible.
There must be enough galaxies formed in the simulation volume to enable us to bin them in a multi-dimensional space at each redshift, not only by reionization redshift, but also by magnitude over a wide range, from as faint as $-11$ to as bright as $-23$.
For example, in order to fit our UVLF to Schechter functions as in Table~\ref{tab:comp}, we must have a large enough sample of galaxies in magnitude bins brighter than $\sim -21$ to model the exponential cut-off without much noise, while also maintaining fine bin-spacing.
For our desired bin size of $\Delta M_\text{UV} = 0.5$, CoDa~II contains $\sim80$ galaxies at $M_\text{UV} = -21$ at $z=6$.
A simulation with half the box size per dimension would, therefore, contain only $\sim10$, and would have almost no galaxies in bins brighter than $-22$.
Thus, apart from their lack of realism in simulating reionization, simulations of smaller volume would under-sample the bright end.  While the number of galaxies is much higher at the faint end, it is also necessary to have the high mass-resolution of CoDa II in order to represent this number and their luminosities faithfully enough to establish the flattening of the UVLF relative to the bright end, as an effect of the suppression of low-mass halos due to reionization. 

In Fig.~\ref{fig:ngtmag}, we plot the cumulative number density of CoDa II galaxies brighter than a given magnitude, again split into $z_\text{re}$ bins. 
The global line in this figure is used for abundance matching to the CoDa II HMF in \S\ref{sec:cv}.

\begin{figure*}
    \centering
    \includegraphics[width=.95\textwidth]{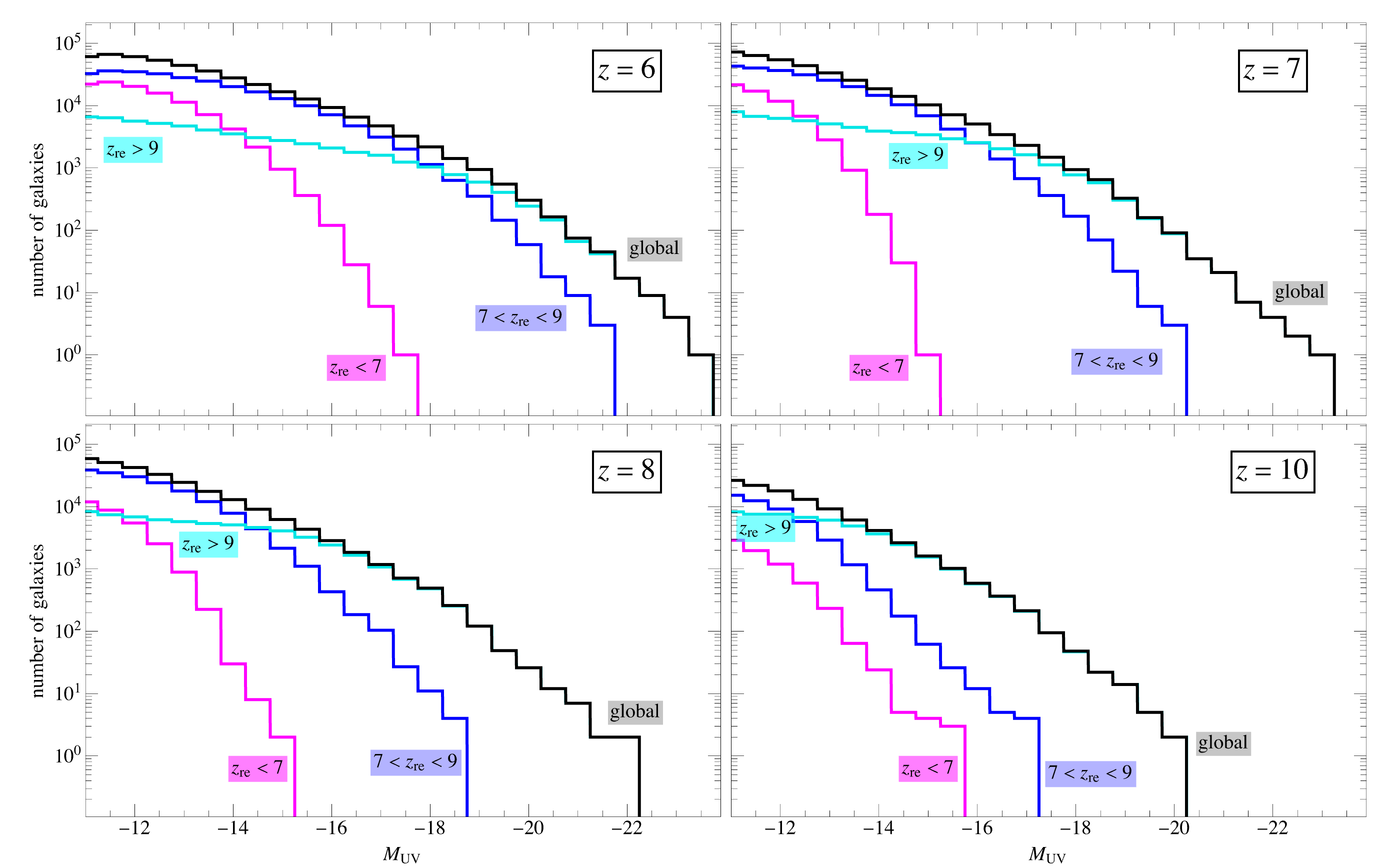}
    \caption{  Histogram of CoDa II galaxies as a function of magnitude for the same $z_\text{re}$ bins and redshifts as previous figures.}
    \label{fig:UVLFnum}
\end{figure*}

\begin{figure*}
    \centering
    \includegraphics[width=.95\textwidth]{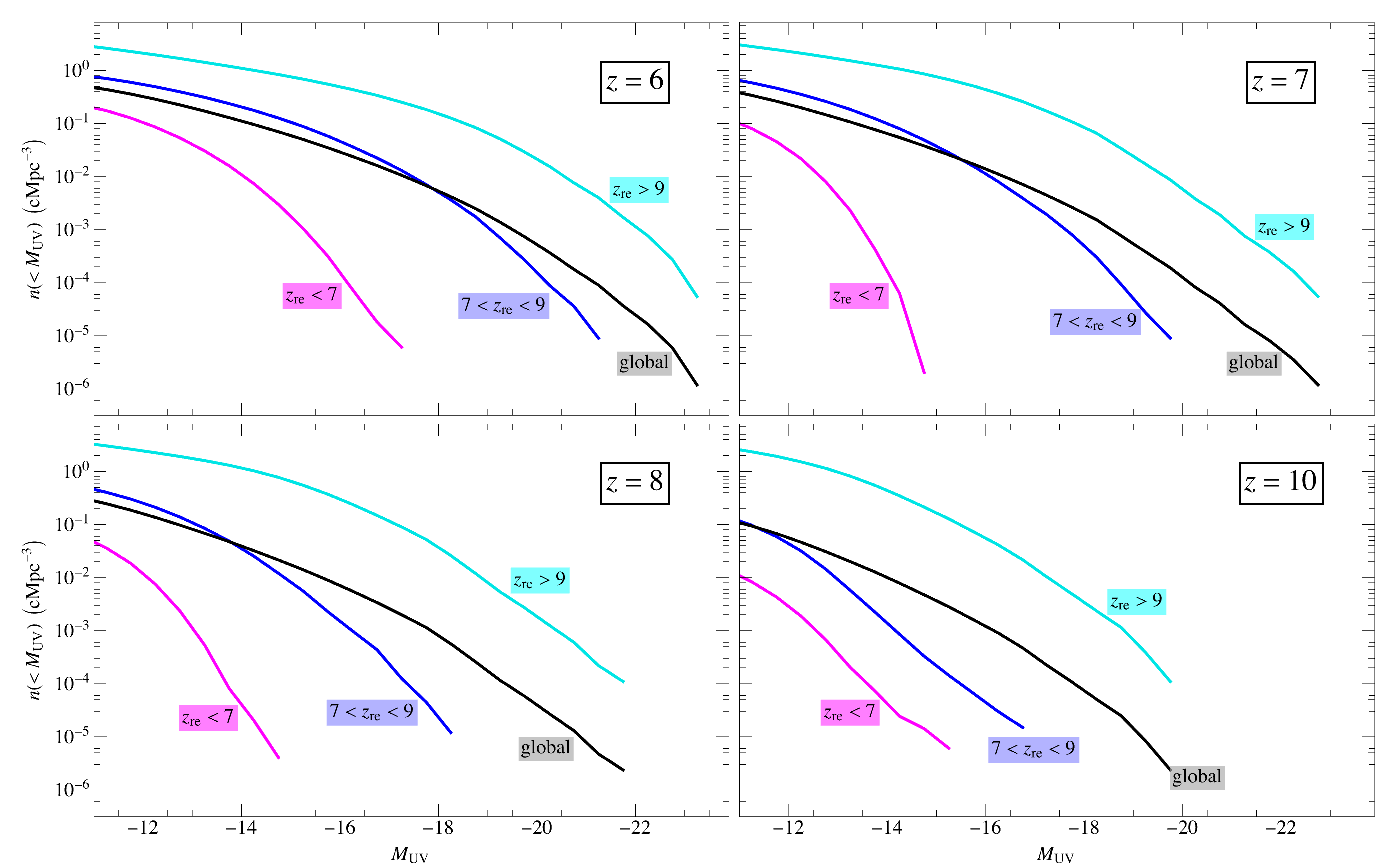}
    \caption{  Cumulative number density of CoDa II galaxies brighter than a given magnitude for the same $z_\text{re}$ bins and redshifts as previous figures.}
    \label{fig:ngtmag}
\end{figure*}

}
%%%%%%%%%%%%%%%%%%%%%%%%%%%%%%%%%%%%%%%%%%%%%%%%%%

% Don't change these lines
\bsp	% typesetting comment
\label{lastpage}
\end{document}